\documentclass[fleqn,usenatbib,useAMS]{mnras}
 \usepackage{graphicx}
 \usepackage{amsmath}
 \usepackage{amssymb}
 \usepackage[T1]{fontenc}
 \usepackage{ae,aecompl}
 \usepackage{txfonts}
 \usepackage{enumerate}

 \title[Asymmetries in extreme trans-Neptunian space]
       {Peculiar orbits and asymmetries in extreme trans-Neptunian space}

 \author[C. de la Fuente Marcos and R. de la Fuente Marcos]
        {C.~de~la~Fuente~Marcos$^{1}$\thanks{E-mail: nbplanet@ucm.es}
         and
         R. de la Fuente Marcos$^{2}$ \\
         $^1$Universidad Complutense de Madrid,
             Ciudad Universitaria, E-28040 Madrid, Spain \\
         $^2$AEGORA Research Group,
             Facultad de Ciencias Matem\'aticas,
             Universidad Complutense de Madrid,
             Ciudad Universitaria, E-28040 Madrid, Spain}
 \date{Accepted 2021 June 12.
       Received 2021 June 10;
       in original form 2021 January 5}
 \pubyear{2021}
 
\begin{document}
  \label{firstpage}
  \pagerange{\pageref{firstpage}--\pageref{lastpage}}
  \maketitle

  \begin{abstract}
     It is still an open question how the Solar system is structured beyond 
     100~au from the Sun. Our understanding of this vast region remains very 
     limited and only recently we have become aware of the existence there of 
     a group of enigmatic bodies known as the extreme trans-Neptunian objects 
     (ETNOs) that have large orbits with perihelia beyond the orbit of 
     Neptune. Four ETNOs ---Sedna, Leleakuhonua, 2012~VP$_{113}$, and 
     2013~SY$_{99}$--- have perihelia beyond 50~au. The study of the ETNOs 
     may provide much needed information on how this remote region is 
     organized. Here, we apply machine-learning techniques to the sample of 
     40 known ETNOs to identify statistically significant clusters that may 
     signal the presence of true dynamical groupings and study the 
     distribution of the mutual nodal distances of the known ETNOs that 
     measure how close two orbits can get to each other. Machine-learning 
     techniques show that the known ETNOs may belong to four different 
     populations. Results from the analysis of the distribution of nodal 
     distances show that 41 per cent of the known ETNOs have at least one 
     mutual nodal distance smaller than 1.45~au (1st percentile of the 
     distribution), perhaps hinting at past interactions. In this context, 
     the peculiar pair of ETNOs made of 505478 (2013~UT$_{15}$) and 
     2016~SG$_{58}$ has a mutual ascending nodal distance of 1.35~au at 
     339~au from the Sun. In addition, the known ETNOs exhibit a highly 
     statistically significant asymmetry between the distributions of object 
     pairs with small ascending and descending nodal distances that might be 
     indicative of a response to external perturbations.
  \end{abstract}

  \begin{keywords}
     methods: statistical -- celestial mechanics --
     minor planets, asteroids: general -- Kuiper belt: general -- 
     Oort Cloud.
  \end{keywords}

  \section{Introduction}
     Our degree of understanding of the structure of the Solar system beyond 100~au from the Sun remains very limited. In order for this 
     situation to change, we have to find and study objects with orbits not only much larger than those of typical members of the 
     trans-Neptunian or Kuiper belt, but also with perihelia increasingly farther from Neptune so their trajectories are only weakly 
     affected by the gravitational pull of the giant planets. The first object fitting into these rather general requirements, 148209 
     (2000~CR$_{105}$), was discovered in February 2000 at Lowell Observatory during a survey of the trans-Neptunian belt 
     \citep{2000DPS....32.2001M,2001MPEC....F...42G}, signalling the presence of an extended scattered structure well beyond Pluto. 

     A trickle of related discoveries has continued throughout the 21st century, in spite of this being an intrinsically challenging task.
     Major milestones in this continuing effort have been the discoveries of 90377 Sedna (2003~VB$_{12}$) in 2003 
     \citep{2004ApJ...617..645B} and of 2012~VP$_{113}$ in 2012 \citep{2014Natur.507..471T}. Although the distinctive nature of this group 
     of objects relative to that of the already well-understood scattered disc was soon recognized (see e.g. \citealt{2002Icar..157..269G,
     2004Natur.432..598K,2004AJ....128.2564M,2005AJ....129..526S}), the origin and evolution of this population (or populations) is far from 
     well established. 

     Here, we adopt the naming convention featured in \citet{2014Natur.507..471T}: extreme (outer) Solar system bodies have perihelion 
     distances, $q>30$~au and semimajor axes, $a>150$~au. These boundaries in $q$ and $a$ are dynamically motivated: $q>30$~au implies that 
     the ETNOs cannot experience close encounters with Neptune and $a>150$~au makes a resonant engagement with Neptune unlikely (see e.g. 
     \citealt{2021arXiv210501065C}). Such a group of minor bodies was designated as exterior trans-Neptunian objects or ETNOs by 
     \citet{2004A&A...428..673R} and extreme trans-Neptunian objects (also ETNOs) by \citet{2014MNRAS.443L..59D}. In the following, we will 
     use the term `extreme trans-Neptunian objects' (ETNOs) to refer to Trujillo and Sheppard's extreme outer Solar system bodies (those 
     with $a>150$~au and $q>30$~au). 

     The ETNOs are unlikely to be captured in distant 1:$N$ mean-motion resonances with Neptune: the farthest ones with known objects in 
     them are 1:9 at $a$$\sim$130~au with 2007~TC$_{434}$ and 2015~KE$_{172}$ \citep{2018AJ....155..260V}, and probably 1:10 at 
     $a$$\sim$140~au with 533563 (2014~JW$_{80}$) and 1:11 at $a$$\sim$150~au with 543735 (2014~OS$_{394}$) as pointed out by 
     \citet{2021arXiv210501065C}. Resonant objects in the 1:12 mean-motion resonance with Neptune at $a$$\sim$157~au and beyond may not 
     exist although \citet{2021arXiv210501065C} showed reasonable numbers of captures as far out as the 1:14, see also the discussion in 
     \citet{2006Icar..184...29G} regarding the 1:18, 1:19 and 1:20 mean-motion resonances with 
     Neptune.\footnote{\href{http://www.fisica.edu.uy/~gallardo/atlas/2020/atlas50to225i10col.png}
     {http://www.fisica.edu.uy/~gallardo/atlas/2020/atlas50to225i10col.png}}$^{,}$\footnote{\href{http://www.fisica.edu.uy/~gallardo/atlas/2020/atlas0to230.png}
     {http://www.fisica.edu.uy/~gallardo/atlas/2020/atlas0to230.png}} The topic of Neptune's resonances in the scattered disc has already 
     been extensively studied and it is relatively well understood within the context of the known Solar system (see e.g. 
     \citealt{2007Icar..192..238L,2016AJ....152..133K,2016ApJ...827L..35N,2017CeMDA.127..477S,2019CeMDA.131...39L,2020CeMDA.132....9G}).
     However, the issue of capture of small bodies in distant mean-motion resonances with Neptune remains an open question. 
     \citet{2021DDA....5230501V} have recently explored the 1:$N$, 2:$N$ and 3:$N$ mean-motion resonances with Neptune out to $a$=550~au at 
     a wide range of perihelion distances ($q$=33--60~au) within the context of the restricted three-body problem to find out that at large 
     $a$, the surviving libration zones of Neptune's strongest resonances are generally wider for larger $q$ due to less crowding from 
     weaker neighbouring resonances, although the sticking times tend to be shorter as one moves further out (see e.g. 
     \citealt{2006Icar..184...29G}). 

     The ETNOs discovered so far (see Table~\ref{etnosB}) move in rather elongated orbits and have perihelion distances beyond 30~au; these 
     unusual properties make them particularly hard to find. In fact, the ETNOs have only been found at perihelion or very near it, leading 
     to a very distinctive pattern regarding how orbital properties relate to discovery locations: their equatorial coordinates at discovery 
     time strongly constraint the orientations of their orbits in space. Orbits are defined by the values of semimajor axis (that controls 
     orbital size), eccentricity, $e$ (that controls shape), and those of the angular elements ---inclination, $i$, longitude of the 
     ascending node, $\Omega$, and argument of perihelion, $\omega$--- that control the orientation of the orbit in space. 
%
%
     \begin{table*}
        \centering
        \fontsize{8}{11pt}\selectfont
        \tabcolsep 0.10truecm
        \caption{Barycentric orbital elements, 1$\sigma$ uncertainties (if known), results of the clustering analysis presented in 
                 Section~3 (cluster number), and number of mutual nodal distances ($X$) below the 1st percentil of ${\Delta}_{+}$, 1.450~au. 
                 The orbit determinations have been computed at epoch JD 2459200.5 that corresponds to 00:00:00.000 TDB on 2020 Dec 17 
                 (J2000.0 ecliptic and equinox). The current orbit determination of 2020~KV$_{11}$ does not include uncertainties. Input 
                 data source: JPL's SBDB. 
                }
        \begin{tabular}{lrrrrrrr}
          \hline
             Object                           & $a_{\rm b}$ (au)   & $e_{\rm b}$         & $i_{\rm b}$ (\degr)    & $\Omega_{\rm b}$ (\degr) & 
                                                 $\omega_{\rm b}$ (\degr) & cluster & $X$ \\
          \hline
                      82158 (2001~FP$_{185}$) &  215.55$\pm$0.04   & 0.84110$\pm$0.00003 & 30.80030$\pm$0.00003   & 179.35849$\pm$0.00004    & 
                                                   6.8745$\pm$0.0005      & 0       & 0   \\
                     148209 (2000~CR$_{105}$) &  221.9$\pm$0.6     & 0.8012$\pm$0.0006   & 22.7558$\pm$0.0006     & 128.2858$\pm$0.0003      & 
                                                 316.691$\pm$0.012        & 0       & 3   \\
            474640 Alicanto (2004~VN$_{112}$) &  328$\pm$2         & 0.8556$\pm$0.0007   & 25.5480$\pm$0.0003     &  66.0222$\pm$0.0004      & 
                                                 326.988$\pm$0.009        & 0       & 3   \\
                     496315 (2013~GP$_{136}$) &  150.2$\pm$0.2     & 0.7269$\pm$0.0004   & 33.5389$\pm$0.0006     & 210.72725$\pm$0.00011    &  
                                                  42.58$\pm$0.04          & 0       & 1   \\
                     508338 (2015~SO$_{20}$)  &  164.78$\pm$0.02   & 0.79871$\pm$0.00002 & 23.41068$\pm$0.00003   &  33.63414$\pm$0.00002    & 
                                                 354.8271$\pm$0.0005      & 0       & 0   \\       
                             2013~RF$_{98}$   &  365$\pm$14        & 0.901$\pm$0.004     & 29.538$\pm$0.003       &  67.636$\pm$0.005        & 
                                                 311.7$\pm$0.7            & 0       & 1   \\
                             2013~UH$_{15}$   &  174$\pm$9         & 0.799$\pm$0.011     & 26.080$\pm$0.006       & 176.542$\pm$0.007        & 
                                                 282.9$\pm$0.3            & 0       & 0   \\ 
                             2014~WB$_{556}$  &  280$\pm$3         & 0.8475$\pm$0.0013   & 24.1575$\pm$0.0002     & 114.891$\pm$0.003        & 
                                                 235.33$\pm$0.06          & 0       & 1   \\
                             2015~BP$_{519}$  &  449$\pm$9         & 0.921$\pm$0.002     & 54.11068$\pm$0.00010   & 135.213$\pm$0.002        & 
                                                 348.06$\pm$0.03          & 0       & 0   \\ 
                             2015~KH$_{163}$  &  153.0$\pm$0.6     & 0.7390$\pm$0.0011   & 27.1377$\pm$0.0014     &  67.5728$\pm$0.0006      &
                                                 230.82$\pm$0.05          & 0       & 1   \\
                             2016~TP$_{120}$  &  175$\pm$25        & 0.77$\pm$0.04       & 32.6395$\pm$0.0006     & 126.73$\pm$0.03          & 
                                                 351.0$\pm$0.4            & 0       & 0   \\
                     523622 (2007~TG$_{422}$) &  502.5$\pm$0.3     & 0.92923$\pm$0.00003 & 18.59537$\pm$0.00003   & 112.91052$\pm$0.00012    & 
                                                 285.6641$\pm$0.0009      & 1       & 1   \\
        541132 Leleakuhonua (2015~TG$_{387}$) & 1090$\pm$126       & 0.940$\pm$0.006     & 11.6714$\pm$0.0006     & 300.995$\pm$0.007        & 
                                                 118.0$\pm$0.3            & 1       & 0   \\
                             2010~GB$_{174}$  &  349$\pm$7         & 0.861$\pm$0.003     & 21.563$\pm$0.002       & 130.716$\pm$0.008        & 
                                                 347.27$\pm$0.08          & 1       & 0   \\ 
                             2013~FT$_{28}$   &  292$\pm$2         & 0.8509$\pm$0.0008   & 17.3754$\pm$0.0014     & 217.722$\pm$0.002        & 
                                                  40.65$\pm$0.05          & 1       & 1   \\
                             2013~RA$_{109}$  &  463$\pm$2         & 0.9006$\pm$0.0004   & 12.39970$\pm$0.00008   & 104.800$\pm$0.004        & 
                                                 262.92$\pm$0.02          & 1       & 0   \\
                             2013~SY$_{99}$   &  736$\pm$28        & 0.932$\pm$0.002     &  4.2253$\pm$0.0012     &  29.510$\pm$0.005        & 
                                                  32.18$\pm$0.11          & 1       & 0   \\ 
                             2013~SL$_{102}$  &  314.4$\pm$0.7     & 0.8787$\pm$0.0003   &  6.50492$\pm$0.00008   &  94.731$\pm$0.006        & 
                                                265.50$\pm$0.06           & 1       & 2   \\
                             2014~FE$_{72}$   & 1621$\pm$445       & 0.978$\pm$0.009     & 20.633$\pm$0.003       & 336.830$\pm$0.005        & 
                                                 133.95$\pm$0.06          & 1       & 0   \\
                             2014~SR$_{349}$  &  299$\pm$21        & 0.841$\pm$0.011     & 17.968$\pm$0.002       &  34.884$\pm$0.015        & 
                                                 341.3$\pm$0.6            & 1       & 0   \\
                             2015~GT$_{50}$   &  311$\pm$3         & 0.8766$\pm$0.0011   &  8.7950$\pm$0.0012     &  46.064$\pm$0.003        & 
                                                 128.99$\pm$0.11          & 1       & 0   \\
                             2015~KG$_{163}$  &  680$\pm$5         & 0.9404$\pm$0.0004   & 13.9942$\pm$0.0012     & 219.103$\pm$0.002        &
                                                  32.10$\pm$0.10          & 1       & 1   \\ 
                             2015~RX$_{245}$  &  423$\pm$5         & 0.8924$\pm$0.0013   & 12.138$\pm$0.002       &   8.6052$\pm$0.0002      & 
                                                 65.12$\pm$0.05           & 1       & 1   \\
                     445473 (2010~VZ$_{98}$)  &  153.434$\pm$0.011 & 0.77612$\pm$0.00002 &  4.510577$\pm$0.000009 & 117.3943$\pm$0.0003      & 
                                                 313.7285$\pm$0.0007      & 2       & 0   \\
                     505478 (2013~UT$_{15}$)  &  200.2$\pm$0.8     & 0.7806$\pm$0.0010   & 10.6521$\pm$0.0010     & 191.9542$\pm$0.0004      & 
                                                 252.13$\pm$0.03          & 2       & 3   \\
                     506479 (2003~HB$_{57}$)  &  159.6$\pm$0.4     & 0.7613$\pm$0.0005   & 15.5004$\pm$0.0003     & 197.8710$\pm$0.0004      &  
                                                  10.837$\pm$0.009        & 2       & 0   \\
                     527603 (2007~VJ$_{305}$) &  192.00$\pm$0.05   & 0.81675$\pm$0.00004 & 11.98365$\pm$0.00007   &  24.38250$\pm$0.00003    & 
                                                 338.3561$\pm$0.0009      & 2       & 1   \\
                             2002~GB$_{32}$   &  206.7$\pm$0.5     & 0.8290$\pm$0.0004   & 14.1921$\pm$0.0002     & 177.0434$\pm$0.0003      & 
                                                  37.048$\pm$0.005        & 2       & 1   \\
                             2003~SS$_{422}$  &  201$\pm$138       & 0.8$\pm$0.2         & 16.78$\pm$0.15         & 151.0$\pm$0.2            & 
                                                 211$\pm$43               & 2       & 0   \\
                             2005~RH$_{52}$   &  153.6$\pm$0.2     & 0.7461$\pm$0.0003   & 20.4456$\pm$0.0004     & 306.1097$\pm$0.0009      & 
                                                  32.513$\pm$0.008        & 2       & 0   \\
                             2015~RY$_{245}$  & 225$\pm$5          & 0.861$\pm$0.003     &  6.0306$\pm$0.0010     & 341.532$\pm$0.006        &
                                                 354.5$\pm$0.2            & 2       & 0   \\
                             2016~GA$_{277}$  &  155$\pm$7         & 0.768$\pm$0.011     & 19.4211$\pm$0.0003     & 112.84$\pm$0.02          & 
                                                 178.52$\pm$0.11          & 2       & 0   \\
                             2016~QU$_{89}$   &  171.5$\pm$0.3     & 0.7944$\pm$0.0004   & 16.9757$\pm$0.0005     & 102.898$\pm$0.004        & 
                                                 303.35$\pm$0.08          & 2       & 1   \\
                             2016~QV$_{89}$   &  171.66$\pm$0.08   & 0.76725$\pm$0.00012 & 21.3874$\pm$0.0002     & 173.2148$\pm$0.0007      &
                                                 281.09$\pm$0.02          & 2       & 0   \\
                             2016~SG$_{58}$   &  232.9$\pm$0.4     & 0.8493$\pm$0.0003   & 13.22090$\pm$0.00005   & 118.978$\pm$0.002        & 
                                                 296.29$\pm$0.04          & 2       & 3   \\       
                             2018~AD$_{39}$   &  166$\pm$7         & 0.767$\pm$0.012     & 19.770$\pm$0.007       & 330.091$\pm$0.014        & 
                                                 49.22$\pm$0.14           & 2       & 0   \\
                             2018~VM$_{35}$   &  261$\pm$64        & 0.83$\pm$0.05       &  8.480$\pm$0.003       & 192.41$\pm$0.06          & 
                                                 303$\pm$3                & 2       & 0   \\
                             2020~KV$_{11}$   &  151.11022838      & 0.77980725          &  4.58080324            & 135.47431392             & 
                                                  52.68155301             & 2       & -   \\
                90377 Sedna (2003~VB$_{12}$)  &  506.4$\pm$0.2     & 0.84954$\pm$0.00005 & 11.928523$\pm$0.000004 & 144.4012$\pm$0.0005      & 
                                                 311.286$\pm$0.003        & 3       & 0   \\
                             2012~VP$_{113}$  &  261.9$\pm$1.5     & 0.693$\pm$0.002     & 24.053$\pm$0.002       &  90.801$\pm$0.006        &
                                                 293.9$\pm$0.4            & 3       & 0   \\
          \hline
        \end{tabular}
        \label{etnosB}
      \end{table*}
%
%

     Figure~1 in \citet{2014MNRAS.443L..59D} shows that the sizes and shapes of the orbits of the ETNOs (panels B and C in the figure) do 
     not depend on the position of the object at discovery, but $i$, $\Omega$ and $\omega$ (panels D, E and F) do. This means that the way 
     observations have been conducted (i.e. where the discoveries are being made) affects the observed distributions of the angular elements 
     (the true distributions are, at this point, unknown). This observational bias implies that ETNOs discovered within a few degrees of 
     each other must have similar orbital orientations. This unfortunate circumstance has led to great controversy regarding the presence 
     (of any kind) of clustering in $\Omega$, $\omega$, and longitude of perihelion, $\varpi=\Omega+\omega$ (see e.g. 
     \citealt{2016AJ....151...22B,2016MNRAS.462.1972D,2017AJ....154...65B,2017AJ....154...50S,2019PhR...805....1B,2019AJ....157..139S,
     2020PSJ.....1...28B,2021ApJ...910L..20B,2021PSJ.....2...59N}). 

     However, the known ETNOs have been discovered by several, unrelated surveys and the samples produced by different surveys with 
     different depths and different pointing histories would result in sets of observables with different sampled distributions, 
     particularly in the case of $i$, $\Omega$ and $\omega$. In addition, the overall distributions of the entire observed sample in the 
     case of $a$ and $e$ are expected to resemble in some way the true distributions. On the other hand, it is often unwarrantedly assumed 
     that the known ETNOs represent a single, monolithic population when the reality could be more complex, as it is in the better-studied 
     trans-Neptunian belt with classical, resonant, scattered and detached objects (see e.g. \citealt{2014AJ....148...55A,
     2018ApJS..236...18B,2020AJ....159..133K}). Whether or not there is a plausible single population of ETNOs remains an open question and 
     here we explore possible answers by applying machine-learning techniques to the sample of 40 known ETNOs to study how are they arranged 
     in orbital parameter space. This paper is organized as follows. In Section~2, we discuss data and methods. Machine-learning techniques, 
     in the form of the $k$-means++ algorithm implemented in the Python library Scikit-learn, are applied in Section~3 to three-dimensional 
     datasets to evaluate the presence of any significant clustering in the sample. Our findings are tested against the distribution of 
     their mutual nodal distances that measures how close two orbits can get to each other in Section~4. Poles and perihelia that define the 
     orientation of the orbits in space are studied in Section~5. Our results are discussed in Section~6 and our conclusions are summarized 
     in Section~7.

  \section{Data and methods}
     Here, we work with publicly available data from Jet Propulsion Laboratory's (JPL) Small-Body Database
     (SBDB)\footnote{\href{https://ssd.jpl.nasa.gov/sbdb.cgi}{https://ssd.jpl.nasa.gov/sbdb.cgi}} and HORIZONS on-line solar system data and
     ephemeris computation service,\footnote{\href{https://ssd.jpl.nasa.gov/?horizons}{https://ssd.jpl.nasa.gov/?horizons}} both provided by
     the Solar System Dynamics Group \citep{2011jsrs.conf...87G,2015IAUGA..2256293G}. The HORIZONS ephemeris system has recently been 
     updated to replace the DE430/431 planetary ephemeris, used since 2013, with the new DE440/441 solution \citep{2021AJ....161..105P} and 
     sixteen most massive small-body perturbers. The new DE440/441 general-purpose planetary solution includes seven additional years of 
     ground and space-based astrometric data, data calibrations, and dynamical model improvements, most significantly involving Jupiter, 
     Saturn, Pluto, and the Kuiper Belt \citep{2021AJ....161..105P}. DE440 covers the years 1550--2650 while DE441 is tuned to cover a time 
     range of $-$13,200 to +17,191 years \citep{2021AJ....161..105P}. The most visible change with this update may be in the ephemerides 
     expressed with respect to the Solar system barycentre. There is a time-varying shift of $\sim$100 km in DE441's barycentre relative to 
     DE431 due to the inclusion of 30 new Kuiper-belt masses, and the Kuiper Belt ring mass \citep{2021AJ....161..105P}.

     In the following, we use barycentric elements because, within the context of the ETNOs, barycentric orbit determinations account better 
     for their changing nature as Jupiter follows its 12~yr orbit around the Sun. Although the values of semimajor axis and eccentricity in 
     barycentric orbit determinations may differ from those of heliocentric ones (for the same object), the ones of the angular elements are 
     nearly the same in both cases \citep{2016MNRAS.462.1972D}. Table~\ref{etnosB} shows the input data used in our analyses that include 
     orbital elements and 1$\sigma$ uncertainties (if known) of 40 known ETNOs. The data (as of 10-June-2021) are referred to epoch 2459200.5 
     Barycentric Dynamical Time (TDB) and they have been retrieved from JPL's SBDB and HORIZONS using tools provided by the Python package 
     Astroquery \citep{2019AJ....157...98G}. Only four known ETNOs have perihelia, $q=a\ (1-e)$, beyond 50~au: 90377 Sedna (2003~VB$_{12}$), 
     541132 Leleakuhonua (2015~TG$_{387}$), 2012~VP$_{113}$, and 2013~SY$_{99}$. The current orbit determination of 2020~KV$_{11}$ does not 
     include 1$\sigma$ uncertainties. The orbit determination of 2003~SS$_{422}$ does include uncertainties, but it is the least precise in 
     the sample shown in Table~\ref{etnosB}.

     In order to analyse the results, we produce histograms using the Matplotlib library \citep{2007CSE.....9...90H} with sets of bins 
     computed using NumPy \citep{2011CSE....13b..22V,2020Natur.585..357H} by applying the Freedman and Diaconis rule \citep{FD81}. The 
     clustering analyses presented here were obtained by applying machine-learning techniques provided by the Python library Scikit-learn 
     \citep{Scikit2011,Scikit2013}; kernel density estimations have been carried out using the Python library SciPy \citep{SciPy2020}.

  \section{Clustering: results from the $k$-means++ algorithm}
     In this section, we have applied the unsupervised machine-learning algorithm $k$-means++ to the data discussed above to evaluate data 
     clustering in barycentric ($a_{\rm b}$, $e_{\rm b}$, $i_{\rm b}$) space (the other two angular elements are not used due to the bias 
     concerns pointed out above). The $k$-means++ algorithm \citep{Kmeans07} performs a centroid-based analysis and it is an improved 
     version of the classical $k$-means algorithm (see e.g. \citealt{Kmeans57,Kmeans67,Kmeans82}). The implementation of $k$-means++ used 
     here is the one included in the method {\tt fit} that is part of the {\tt KMeans} class provided by the Python library Scikit-learn 
     \citep{Scikit2011}. The method {\tt fit\_predict} computes cluster centres and returns cluster labels or indexes. As the threshold to 
     define the final clusters in the dataset, we have used the elbow method to determine the optimal value of clusters, $k$; we invoke the 
     method {\tt fit} with $k$ in the interval (1, 10) and select the value of $k$ that minimizes the sum of the distances of all data 
     points to their respective cluster centres. As part of the data preparation process, we have scaled the dataset using standardisation 
     or Z-score normalisation: found the mean and standard deviation for $a_{\rm b}$, $e_{\rm b}$ and $i_{\rm b}$, subtracted the relevant 
     mean from each value, and then divided by its corresponding standard deviation. This was carried out by applying the method 
     {\tt fit\_transform} that is part of the {\tt StandardScaler} class provided by the Python library Scikit-learn \citep{Scikit2011}. 
     Distance assignment between the objects in our sample assumes a Euclidean metric. 

     \subsection{Trans-Neptunian objects}
\begin{figure*}
       \centering
        \includegraphics[width=0.33\linewidth]{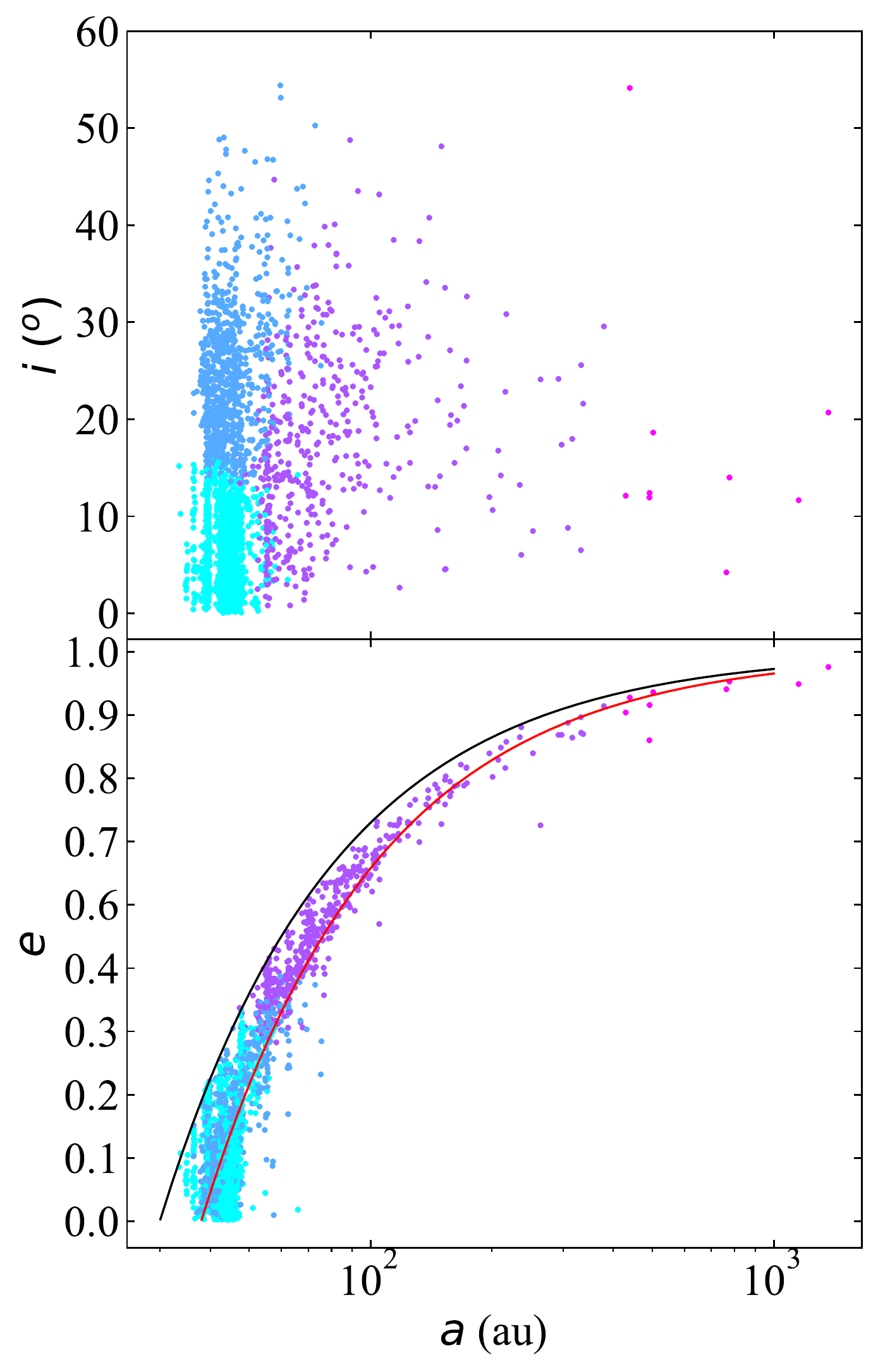}
        \includegraphics[width=0.33\linewidth]{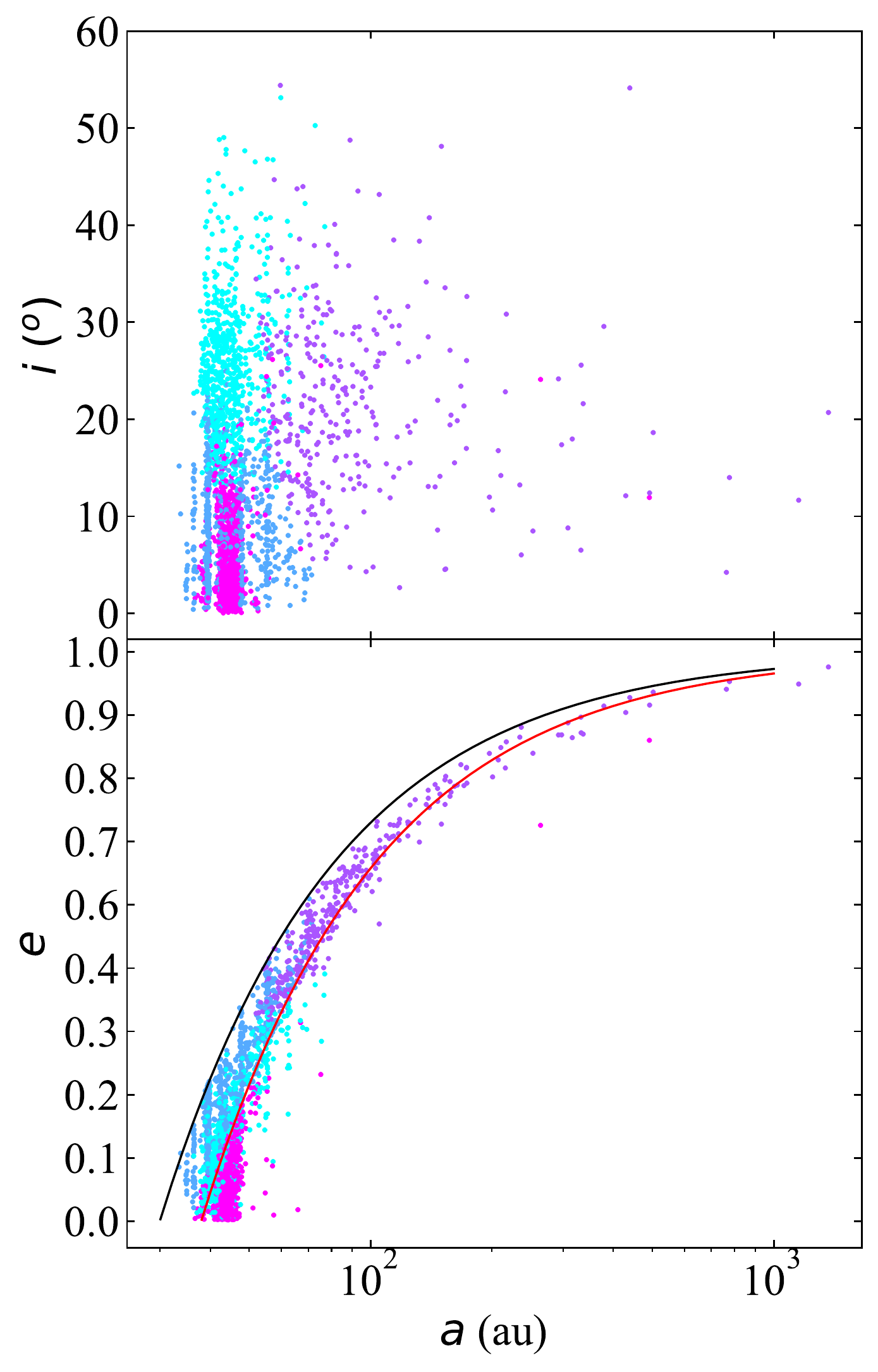}
        \includegraphics[width=0.33\linewidth]{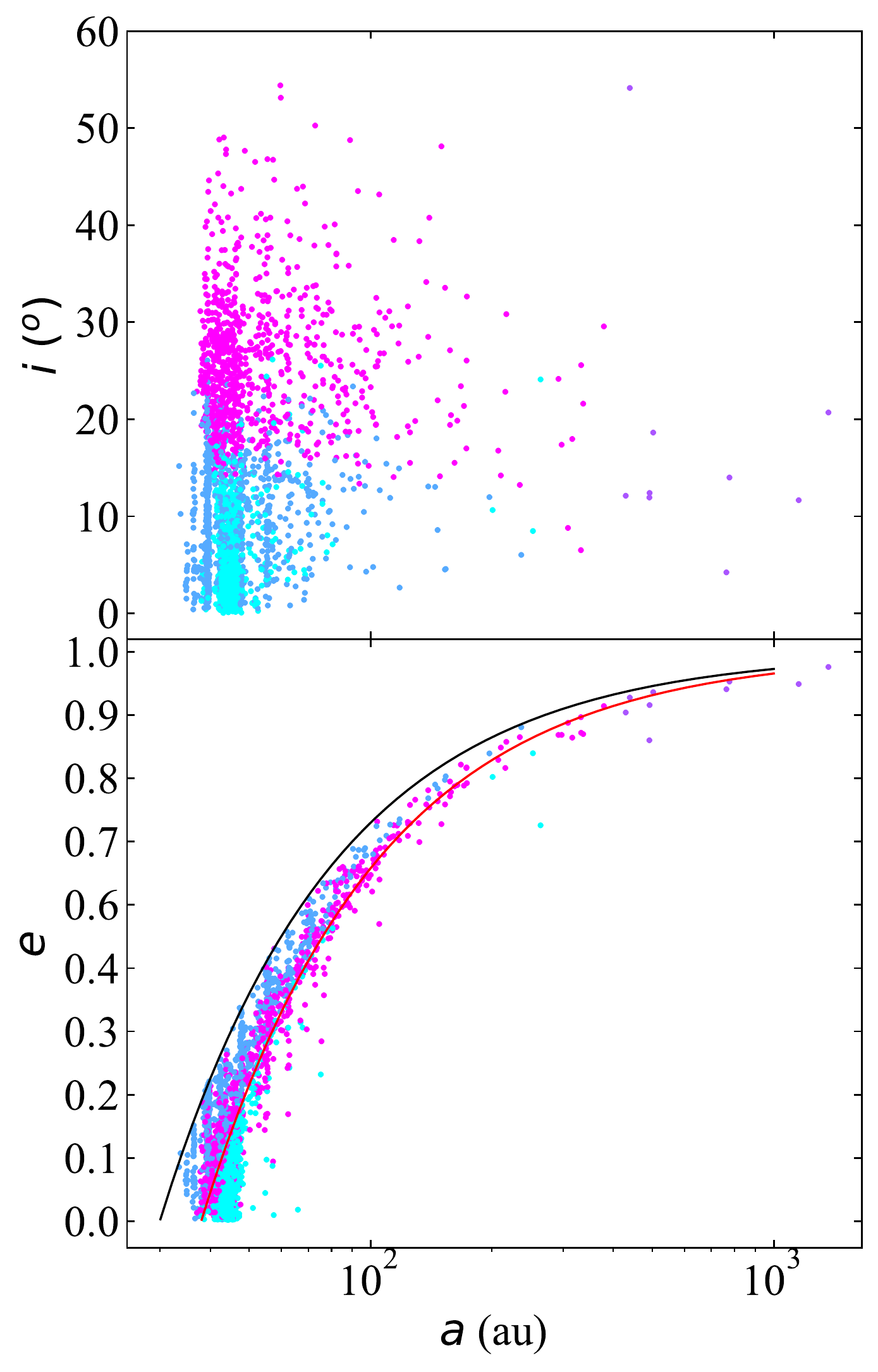}
        \caption{Colour-coded clusters generated by the $k$-means++ algorithm applied to the dataset made of 3119 trans-Neptunian objects 
                 (TNOs). The left-hand side panels correspond to the algorithm being applied to the dataset ($a$, $e$, $i$); the central
                 panels are for ($q$, $e$, $i$); the right--hand side panels are for ($a$, $q$, $i$). The black curve shows $q=30$~au and 
                 the red one $q=38$~au. In the left-hand side panels, the cold (aqua) and hot (azure) components of the classical TNOs are 
                 well separated by the algorithm, scattered and detached TNOs appear mixed (plum), and the nine objects with the largest 
                 values of $a$ appear as a separate group (fuchsia). In the central panels, the cold (fuchsia) and hot (aqua) components of 
                 the classical TNOs and the resonant TNOs (azure) are relatively well identified, scattered and detached TNOs appear mixed 
                 (plum). The right--hand side panels show the cold (aqua) component of the classical TNOs, the resonant TNOs (azure) appear 
                 mixed with scattered and detached TNOs that also mingle with the hot (fuchsia) component, and the nine objects with the 
                 largest values of $a$ appear as a separate group (plum). 
                }
        \label{clustersTNOs}
     \end{figure*}

        In order to evaluate the reliability of the results of the unsupervised machine-learning algorithm $k$-means++ within the context of
        the trans-Neptunian populations, we have applied the sequence of steps described above to the (heliocentric orbital elements) 
        dataset made of 3119 trans-Neptunian objects or TNOs with $q>30$~au. Our approach is essentially different from the one discussed by
        \citet{2020MNRAS.497.1391S}. Known TNOs are classified as classical, resonant, scattered and detached objects (see e.g. 
        \citealt{2014AJ....148...55A,2018ApJS..236...18B,2020AJ....159..133K}). Classical TNOs have $a\in(42, 46)$~au with a cold component 
        with $e\leq0.1$ and $i\leq5\deg$ and a hot one with higher values of $e$ and $i$. Resonant TNOs are trapped in mean-motion 
        resonances with Neptune. Scattered TNOs have $a\in(30, 1000)$~au and $q\in(30, 38)$~au. Detached TNOs have $a\geq50$~au and 
        $q\geq40$~au. Figure \ref{clustersTNOs} shows the results of the application of $k$-means++ and the elbow method to the TNO sample; 
        results are different depending on the orbital parameters used to perform the multiparametric analysis although the three analyses 
        produce four clusters. While unsupervised machine-learning techniques applied to orbital elements are not expected to reproduce 
        precisely the results of classification algorithms based on extensive numerical simulations, they indeed show that four distinct 
        populations may be present in the full TNO sample. Our results are purely statistical, without any input from $N$-body simulations. 
        But, which (tri-parametric) dataset does produce the most consistent results when considering the actual dynamical classes pointed 
        out above?
%
%
%
%

        The left-hand side panels of Fig.~\ref{clustersTNOs} are the result of focusing on ($a$, $e$, $i$) and fail to show the resonant 
        TNOs as a separate dynamical class. The cold (aqua) and hot (azure) components of the classical TNOs are clearly identified albeit 
        also including the resonant TNOs and perhaps others. Scattered and detached TNOs appear mixed (plum). The nine objects with the 
        largest values of $a$ appear as a separate group (fuchsia). The central panels of Fig.~\ref{clustersTNOs} are for ($q$, $e$, $i$). 
        In this case, resonant TNOs (azure) are relatively well identified at $i<20\deg$ and also the cold (fuchsia) and hot (aqua) 
        components of the classical TNOs. Scattered and detached TNOs mingle (plum) as they did when we considered ($a$, $e$, $i$). In our 
        third analysis, the right--hand side panels of Fig.~\ref{clustersTNOs} show the results of focusing on ($a$, $q$, $i$). The cold 
        (aqua) component of the classical TNOs is clearly identified but the hot one appears mixed with scattered and detached TNOs 
        (fuchsia). The resonant TNOs (azure) also appear mixed with scattered and detached TNOs. The nine objects with the largest values 
        of $a$ are identified as a separate group (plum).

        The virtual absence of TNOs moving in low-eccentricity, low-inclination orbits beyond about 50~au and visible in 
        Fig.~\ref{clustersTNOs} was first noticed by \citet{1997ASPC..122..347D}, \citet{1998AJ....116.2042G} and 
        \citet{1998AJ....115.2125J}. \citet{1999AJ....118.1411C} coined the term `Kuiper Cliff' to refer to such a feature and it has been 
        used ever since to describe the unexpected drop-off in the fraction of TNOs following dynamically cold paths. The reality of the 
        Kuiper Cliff has been demonstrated by multiple studies (see e.g. \citealt{2002AJ....124.3430C,2003EM&P...92..113B,
        2003AJ....126..430C,2004AJ....128.1364B,2007AJ....133.1247L}).

        Although the results of the clustering analysis carried out with the $k$-means++ algorithm cannot reproduce the widely recognized
        dynamical classes inhabiting trans-Neptunian orbital parameter space, the analysis based on ($q$, $e$, $i$) is perhaps the one 
        providing the most approximate results. All the analyses fail to separate scattered and detached TNOs properly, which is to be 
        expected as they have disperse distributions, not well suited to be resolved by any type of clustering algorithm.

     \subsection{Extreme trans-Neptunian objects}
\begin{figure*}
       \centering
        \includegraphics[width=0.33\linewidth]{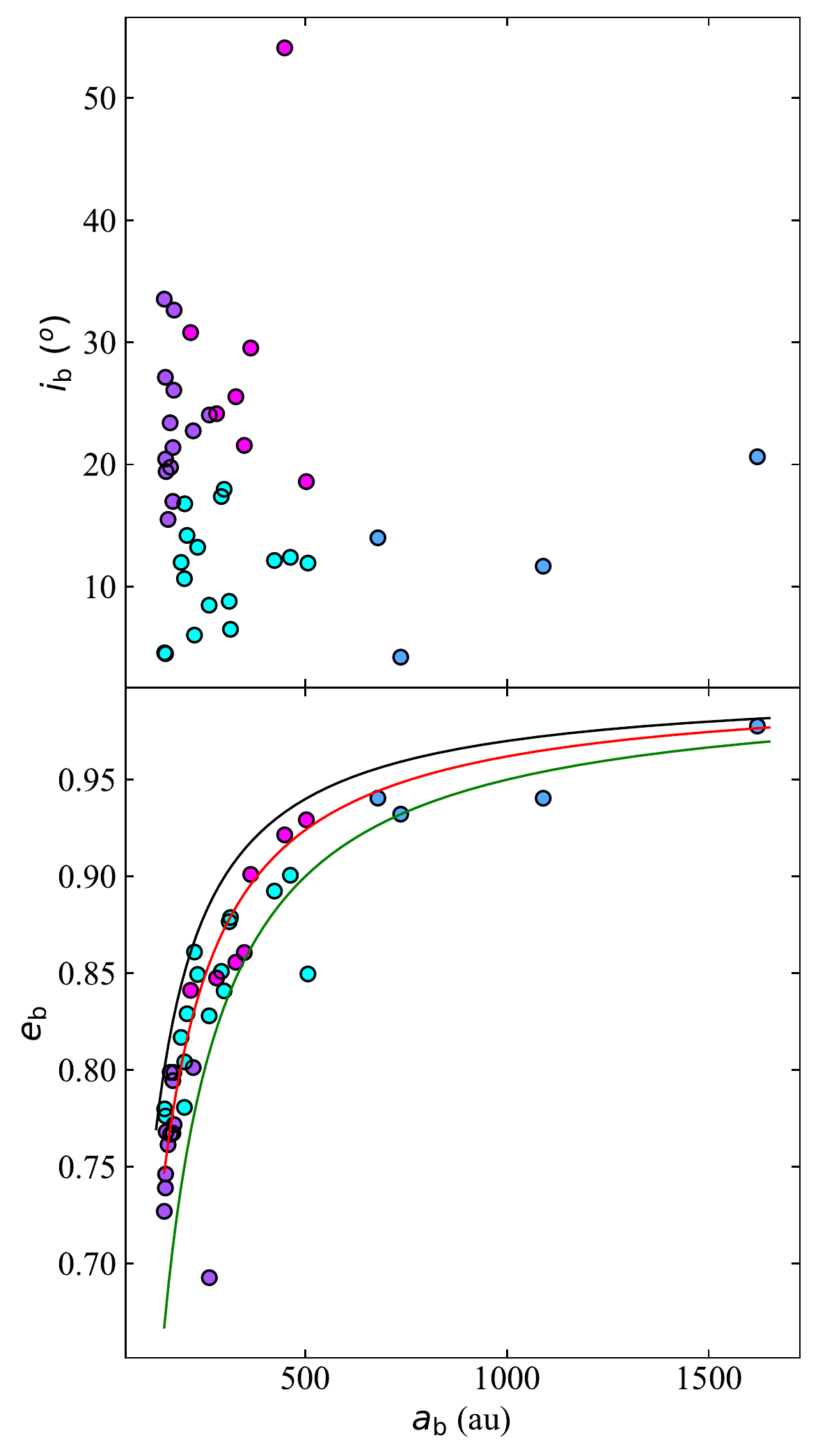}
        \includegraphics[width=0.33\linewidth]{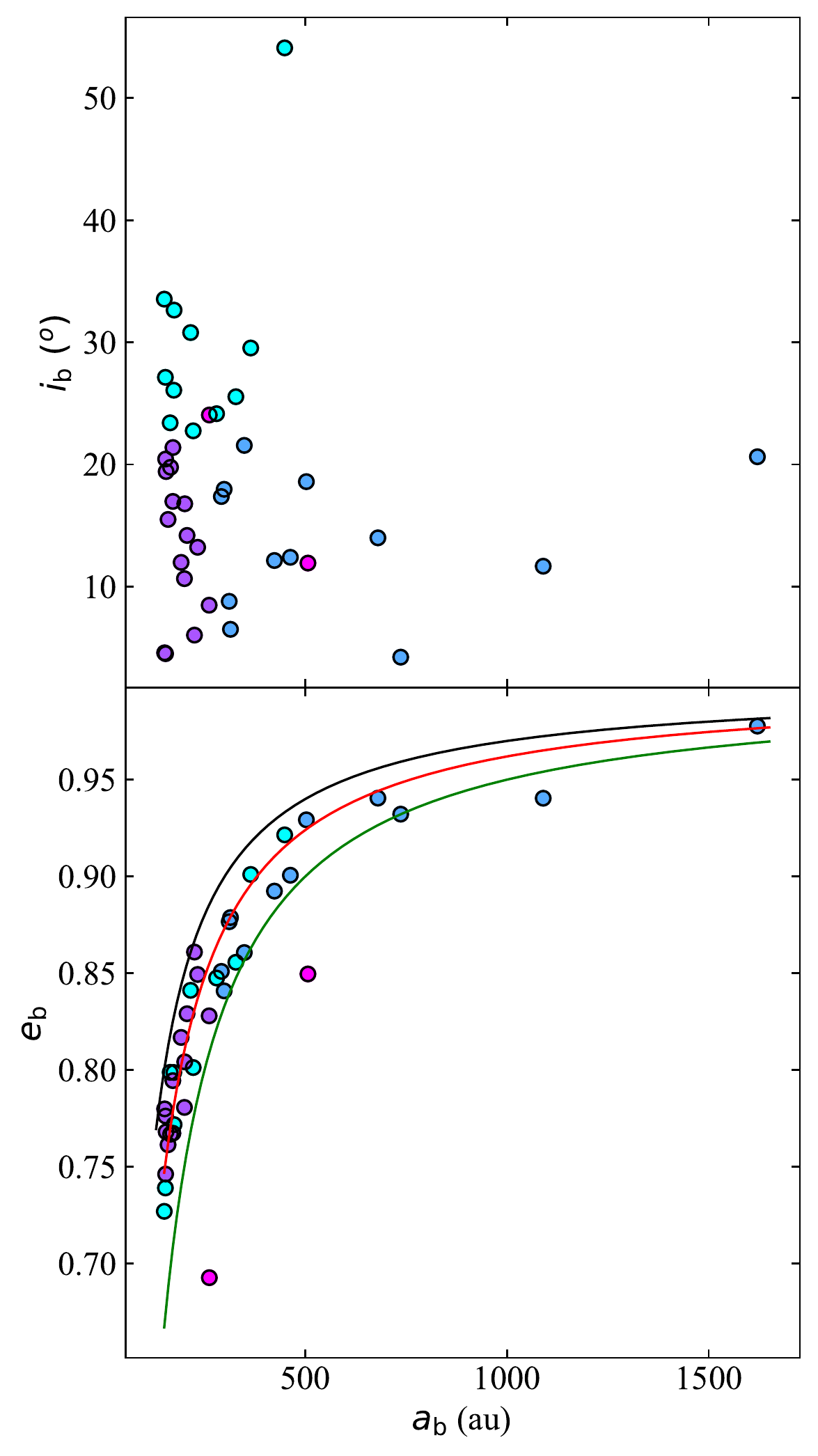}
        \includegraphics[width=0.33\linewidth]{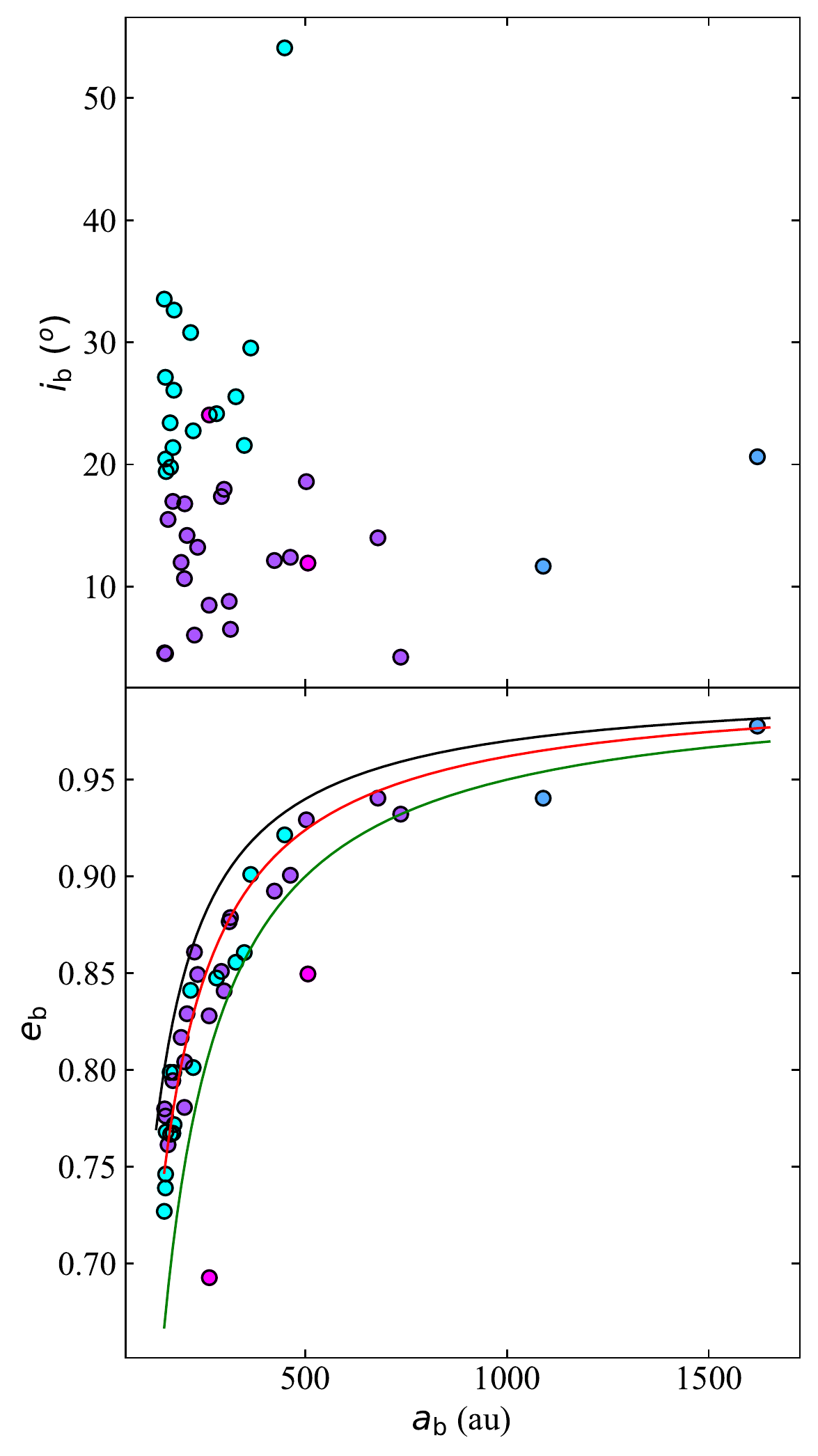}
        \caption{Colour-coded clusters generated by the $k$-means++ algorithm applied to the dataset made of 40 extreme trans-Neptunian 
                 objects (ETNOs). The left-hand side panels correspond to the algorithm being applied to the dataset ($a$, $e$, $i$); the 
                 central panels are for ($q$, $e$, $i$); the right--hand side panels are for ($a$, $q$, $i$), see the text for details. The 
                 black curve shows $q=30$~au, the red one $q=38$~au, and the green one $q=50$~au. Focusing on the clusters from the ($q$, $e$, 
                 $i$) analysis: the aqua-coloured points (11) correspond to objects labelled as `0' in Table~\ref{etnosB}, the azur points
                 (12) correspond to those labelled as `1', the plum points (15) correspond to those labelled as `2', and the fuchsia points 
                 (2) correspond to ETNOs labelled as `3' in the table.
                }
        \label{clustersETNOs}
     \end{figure*}

        When applying the $k$-means++ algorithm and the elbow method to the ETNO dataset, we obtained four clusters that are shown in 
        Fig.~\ref{clustersETNOs}; a consistent result is obtained when heliocentric orbit determinations are used instead of barycentric 
        ones. The central panels of Fig.~\ref{clustersETNOs} are for ($q$, $e$, $i$) that produced the best matching for the full TNO 
        sample in terms of dynamical groupings. In the following, we will focus on the results of this particular clustering analysis.
        The colours in Fig.~\ref{clustersETNOs} are linked to the number labels (cluster indexes) in Table~\ref{etnosB}: 0 in aqua, 1 in 
        azur, 2 in plum, and 3 in fuchsia. The reference object for cluster~0 is 148209 (2000~CR$_{105}$), for cluster~1 is 541132 
        Leleakuhonua (2015~TG$_{387}$), the one for cluster~2 is 445473 (2010~VZ$_{98}$), and the reference object for cluster~3 is 90377 
        Sedna (2003~VB$_{12}$). There is an object, 2015~BP$_{519}$ (in cluster~0, see Fig.~\ref{clustersETNOs}, top panel), that is a 
        statistical outlier in terms of the inclination value, 54\fdg11068$\pm$0\fdg00010 \citep{2018AJ....156...81B,2018RNAAS...2..167D}.
%
%
%
%
     
     In our analysis, the Sednoids (those objects following Sedna-like orbits) emerge as a statistically significant cluster (\#3 in 
     fuchsia, Fig.~\ref{clustersETNOs}, central panels), the smallest of the four found here, with just one member other than Sedna, 
     2012~VP$_{113}$. Although clusters~0 and 2 may be part of the scattered and detached TNO dynamical classes, cluster~1 may represent a 
     separate population, perhaps linked to the so-called inner Oort Cloud \citep{1981AJ.....86.1730H,2001AJ....121.2253L}. Both 
     Leleakuhonua and 2013~SY$_{99}$ are members of cluster~1. Sedna, Leleakuhonua, 2012~VP$_{113}$, and 2013~SY$_{99}$ (barely) are the 
     only known objects with $q>50$~au, i.e. beyond the Kuiper Cliff. The four statistically significant clusters may represent true 
     dynamical groupings, but additional analyses are required before reaching a final conclusion regarding this issue. In any case, 
     statistics alone strongly suggests that the known ETNOs do not belong to a single dynamical group but to a mixture of several, probably 
     four, populations.

  \section{Nodal distances}
\begin{figure}
       \centering
        \includegraphics[width=\linewidth]{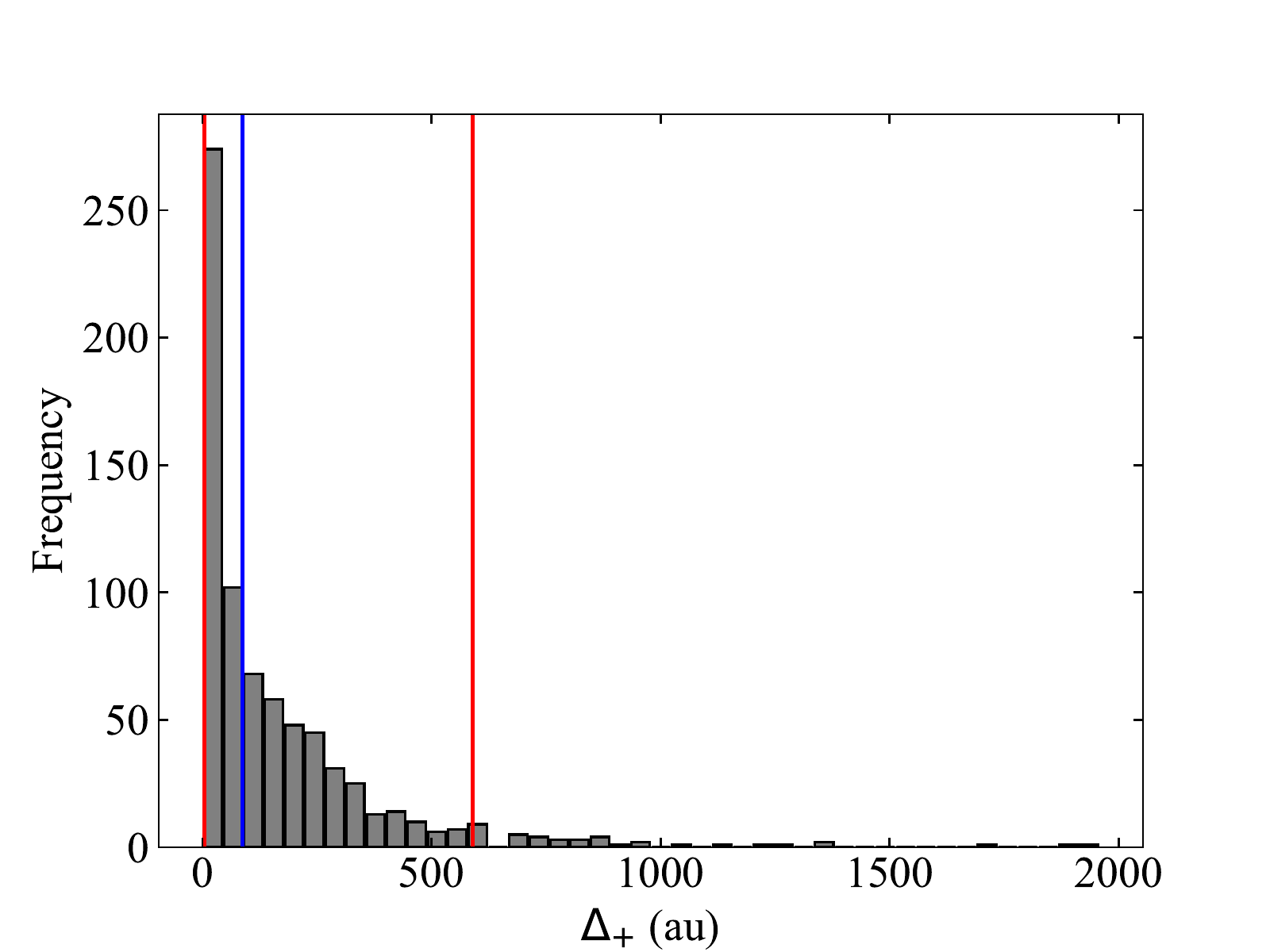}
        \caption{Distribution of mutual nodal distances for the ascending nodes of the sample of 39 ETNOs. The median is shown in blue and 
                 the 5th and 95th percentiles in red.  
                }
        \label{ascnodes}
     \end{figure}

\begin{figure}
       \centering
        \includegraphics[width=\linewidth]{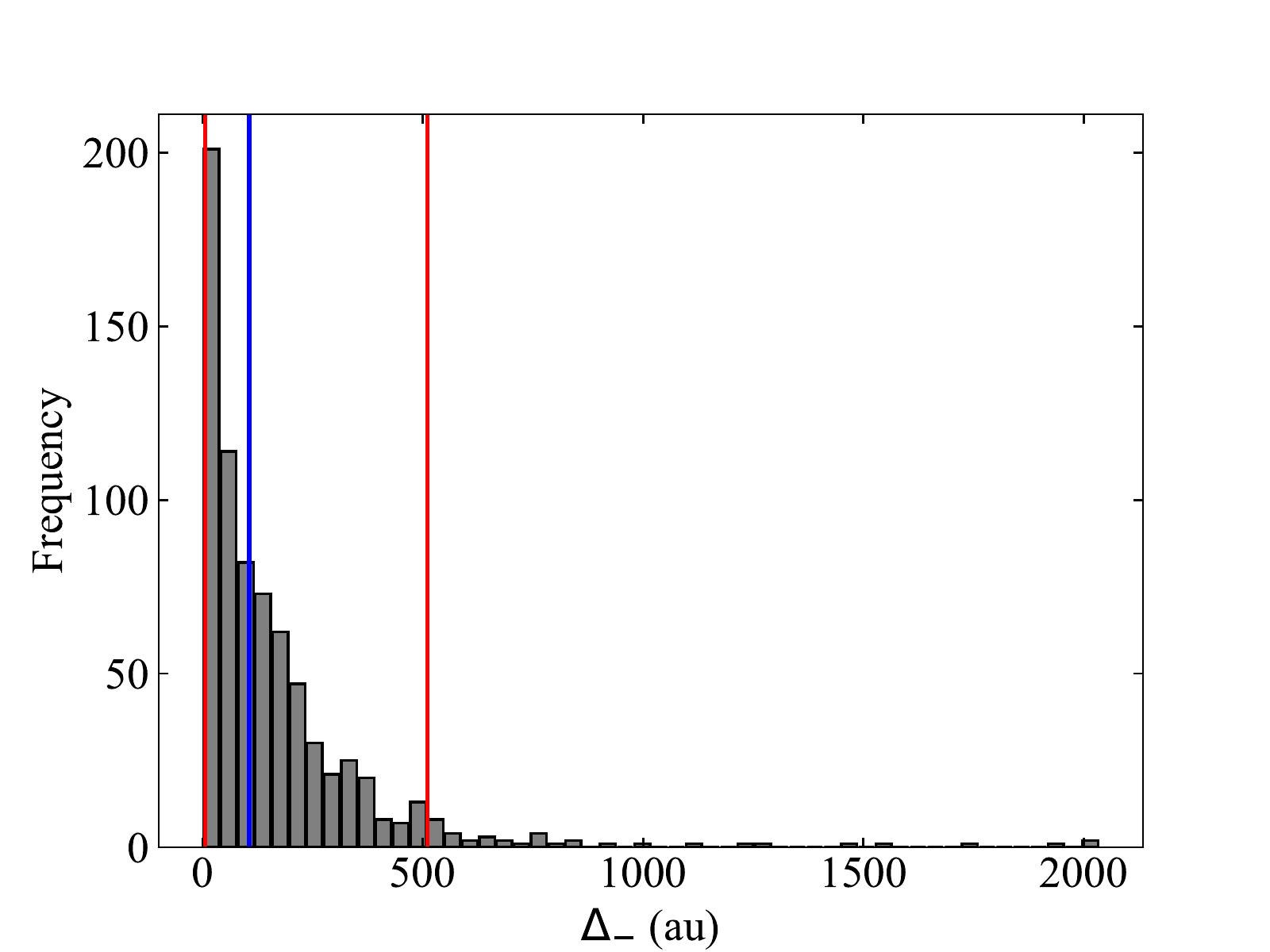}
        \caption{Distribution of mutual nodal distances for the descending nodes of the sample of 39 ETNOs. The median is shown in blue and  
                 the 5th and 95th percentiles in red. 
                }
        \label{desnodes}
     \end{figure}

     If real, the four statistically significant clusters singled out in the previous section must produce consistent structures in other
     distributions of relevant orbital parameters. Here, we focus on the distribution of mutual nodal distances (absolute values) computed 
     as described in Appendix~A using data from Table~\ref{etnosB}. A small mutual nodal distance implies that in absence of protective 
     mechanisms such as mean-motion or secular resonances, the objects may experience close flybys and perhaps collisions. Discarding 
     2020~KV$_{11}$ because its orbit determination is still too uncertain, we have a sample of 39 ETNOs that produced 741 pairs (i.e. 1482 
     in total) of mutual nodal distances, ${\Delta}_{\pm}$ (the results for each pair come from a set of 10$^4$ pairs of virtual ETNOs as 
     described in Appendix~A). The distribution of mutual nodal distances for the ascending nodes is shown in Fig.~\ref{ascnodes} and the 
     one corresponding to the descending nodes is displayed in Fig.~\ref{desnodes}. These distributions have been computed using the data in 
     Table~\ref{etnosB} and considering the uncertainties in the orbit determinations as described in Appendix~A. 
%
%
%
%
%
%
%
%

\begin{figure}
       \centering
        \includegraphics[width=\linewidth]{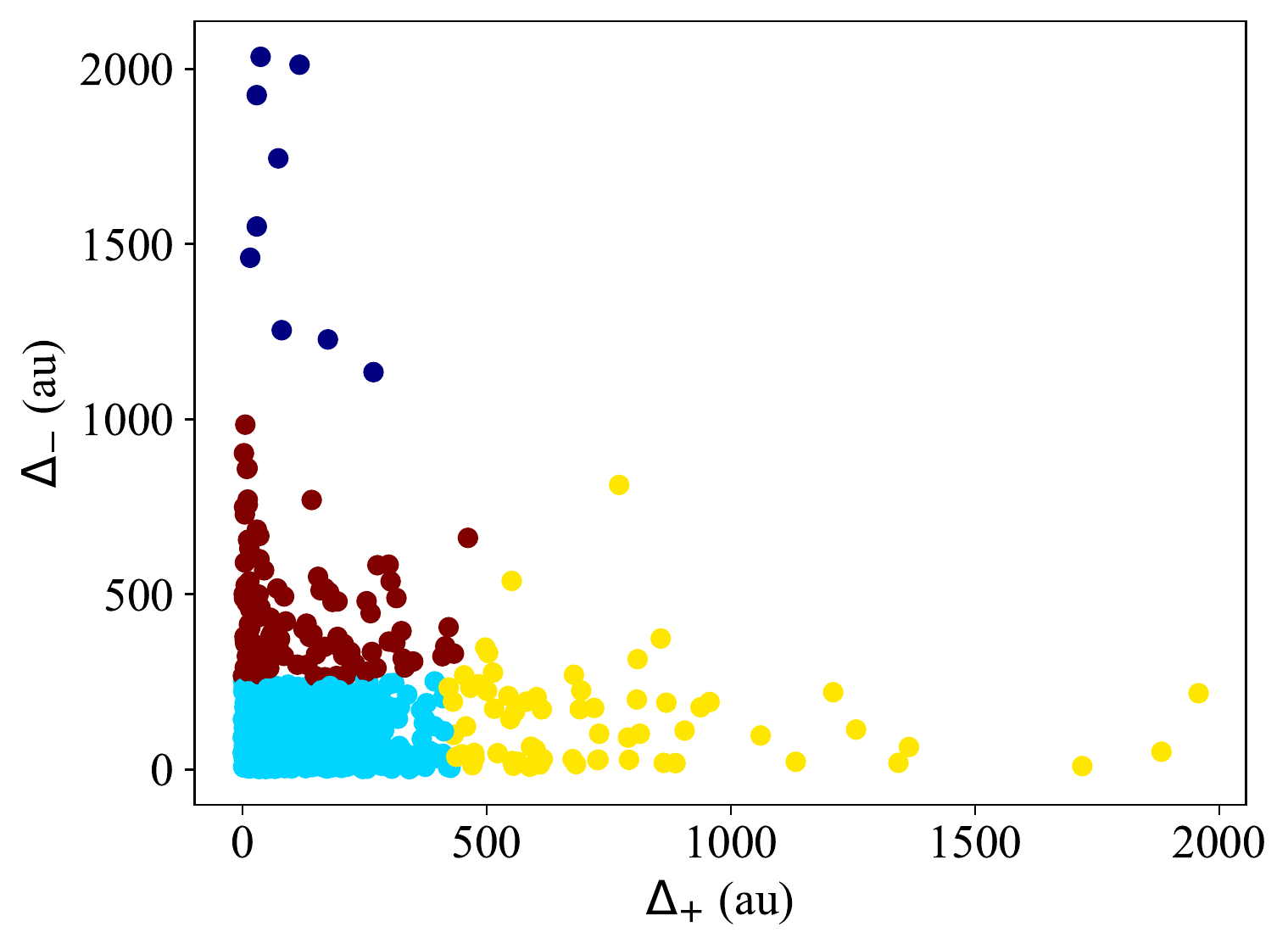}
        \caption{Colour-coded clusters generated by the $k$-means++ algorithm applied to the dataset made of 741 pairs of mutual nodal 
                 distances from the sample of 39 ETNOs.
                }
        \label{nodeclusters}
     \end{figure}

\begin{figure}
       \centering
        \includegraphics[width=\linewidth]{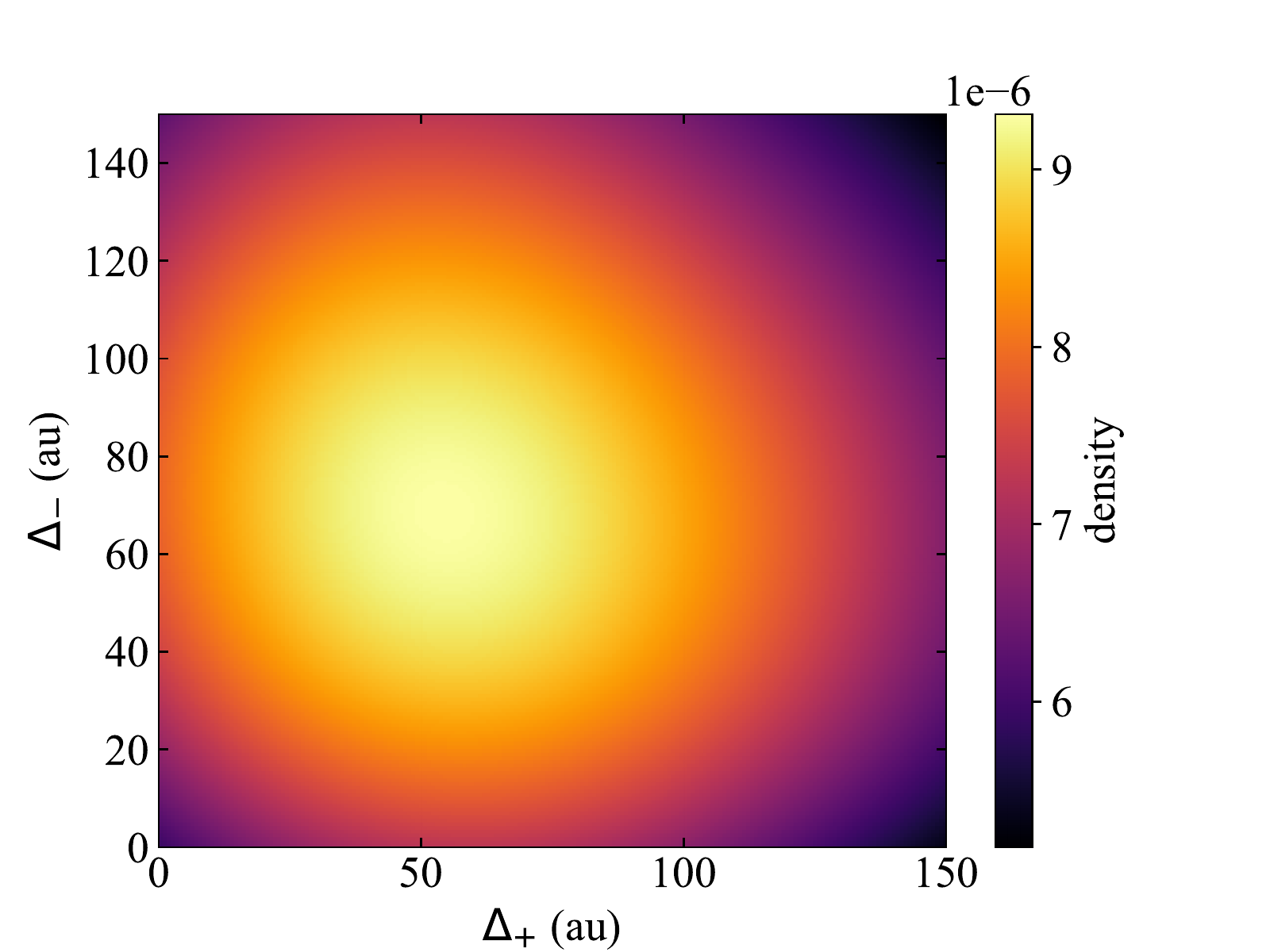}
        \caption{Gaussian kernel density estimation of the same data shown in Fig.~\ref{nodeclusters} but focusing in the subsample with the
                 smallest values of ${\Delta}_{\pm}$ (the region in dark blue in Fig.~\ref{nodeclusters}). 
                }
        \label{nodemap}
     \end{figure}

     Applying the $k$-means++ algorithm to the ${\Delta}_{\pm}$ dataset as described in the previous section, we obtained 
     Fig.~\ref{nodeclusters} that again displays four clusters. The coincidence in the number of clusters is significant as the nodal 
     distance analysis uses the uncertainties in the orbit determinations and includes all the orbital elements, not just ($a_{\rm b}$, 
     $e_{\rm b}$, $i_{\rm b}$). Finding four clusters in terms of nodes means that one of the clusters identified in the previous section is
     not interacting with the other three. The region in aqua in Fig.~\ref{nodeclusters} corresponds to smaller values of ${\Delta}_{\pm}$ 
     linked to orbits that in absence of protective resonant mechanisms may lead to relatively close flybys, the others include pairs of 
     objects whose orbits cannot intersect as they are fully detached. We interpret this result as supportive of the presence of four true 
     dynamical groups in the sample of 39 ETNOs. The Gaussian kernel density estimation in Fig.~\ref{nodemap} shows that there are a number 
     of pairs with unusually low values of the mutual nodal distances considering the large sizes of the orbits involved.
%
%
%
%
%
%
%
%

\begin{figure*}
       \centering
        \includegraphics[width=\linewidth]{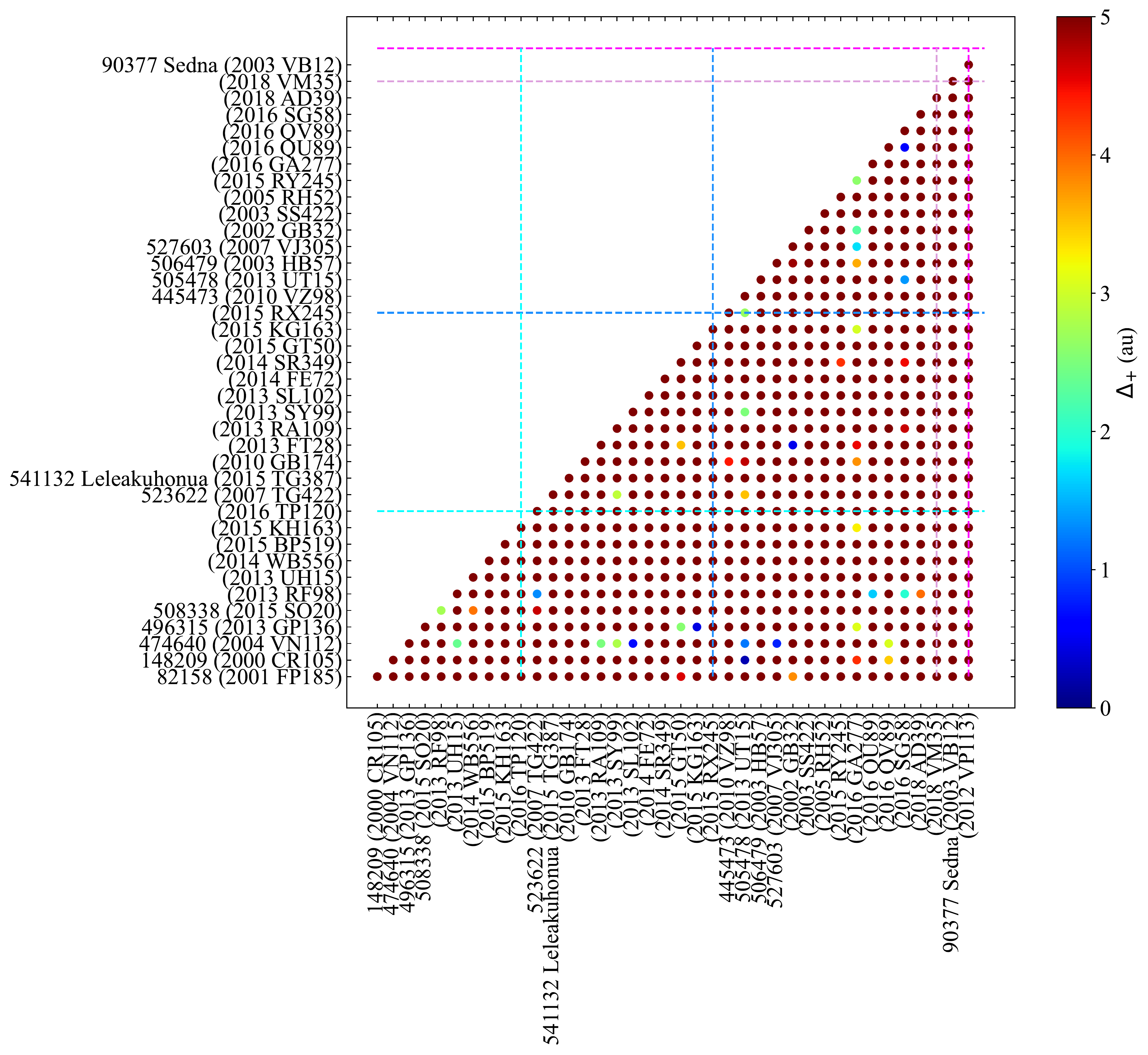}
        \caption{Colour-coded mutual ascending nodal distances for all the 741 pairs. As we are focusing on unusually low values of the 
                 mutual nodal distances, only the range 0--5~au is shown in detail. Values of ${\Delta}_{\pm}>5$~au appear in brown. The 
                 ETNOs have been sorted as they appear in Table~\ref{etnosB} and the statistically significant clusters identified in 
                 Section~3.2 are indicated using dashed horizontal and vertical lines of colours consistent with those in 
                 Fig.~\ref{clustersETNOs}, central panels.
                }
        \label{ascnodesALL}
     \end{figure*}

\begin{figure*}
       \centering
        \includegraphics[width=\linewidth]{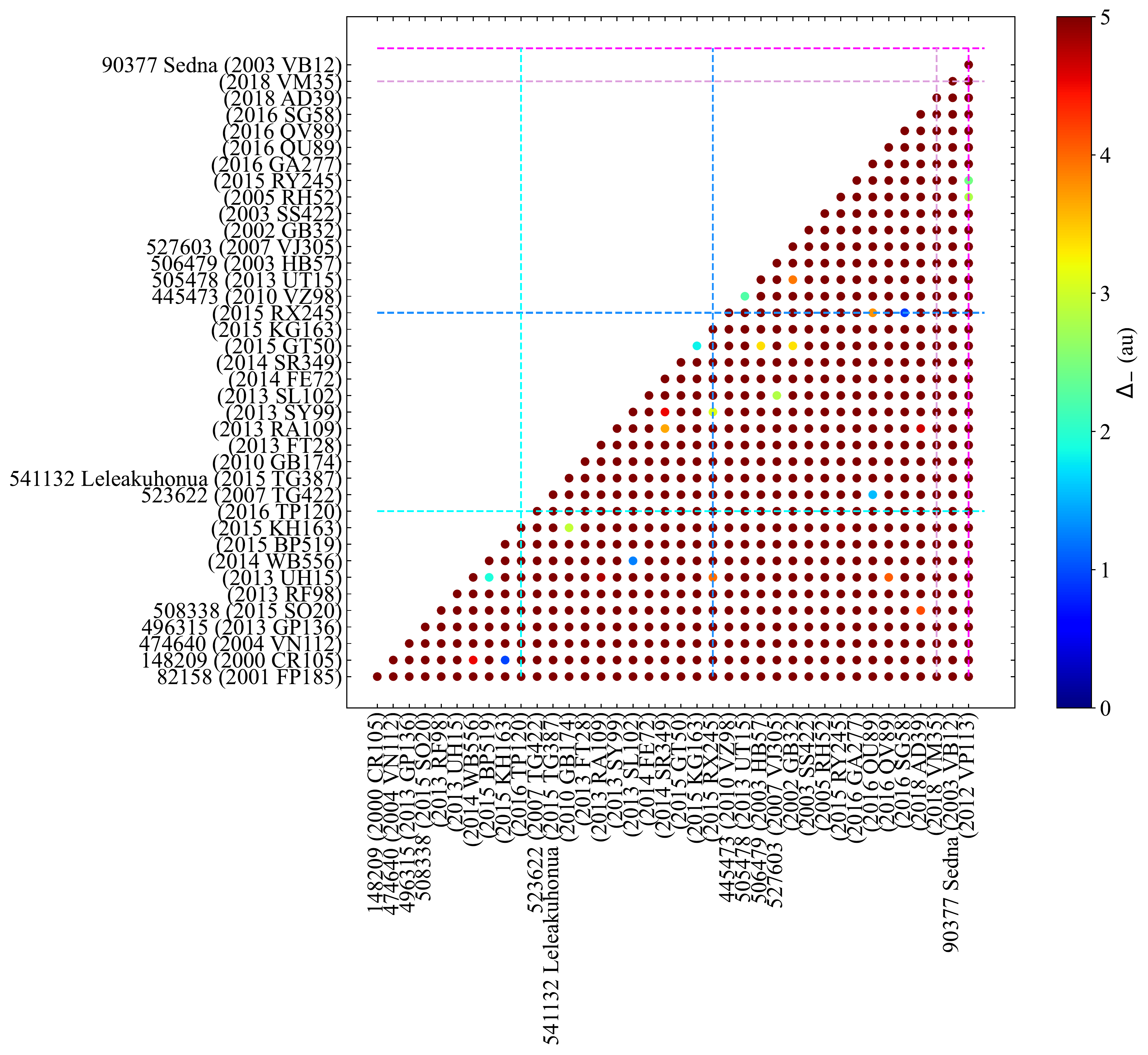}
        \caption{As Fig.~\ref{ascnodesALL} but for the descending nodes.
                }
        \label{desnodesALL}
     \end{figure*}

     This is more clearly seen when considering Figs~\ref{ascnodesALL} and \ref{desnodesALL}, where the nodal distances of each pair appear 
     colour coded. In these figures, the ETNOs have been sorted ($x$-axis from left to right, $y$-axis from bottom to top) as they appear in 
     Table~\ref{etnosB} and the statistically significant clusters identified in Section~3.2 have been indicated using dashed horizontal and 
     vertical lines of colours consistent with those in Fig.~\ref{clustersETNOs}, central panels. Figures~\ref{ascnodesALL} and 
     \ref{desnodesALL} show an obvious asymmetry as there are far more low values of mutual ascending nodal distances (Fig.~\ref{ascnodesALL}) 
     than of the descending ones (Fig.~\ref{desnodesALL}). The figures show the median values of the mutual nodal distances that have been 
     computed as described in Appendix~A, taking into account the uncertainties in Table~\ref{etnosB} by using a Monte Carlo approach. 
     Figures~\ref{ascnodesALL} and \ref{desnodesALL} show that there are two types of ETNOs, those that follow orbits that pass rather close 
     to the paths of other ETNOs and those in trajectories well detached from the rest. ETNOs in cluster 3, the Sednoids, do not pass close 
     to any other known ETNOs. There are no known pairs with low values of both mutual nodal distances. Among the pairs with low values of 
     one mutual nodal distance, only a few are made of ETNOs within the same cluster. Most low values appear in pairs that include members of 
     clusters 0 and 2. Cluster 1 has comparatively less members in pairs with low values of the mutual nodal distances than clusters 0 and 
     2.
%
%
%
%
%
%
%
%

     In order to identify severe outliers, we use the 1st percentile of the distribution (see e.g. \citealt{2012psa..book.....W}) that is 
     1.450$\pm$0.010~au for ${\Delta}_{+}$ and 2.335$\pm$0.014~au for ${\Delta}_{-}$ (the dispersions come from ten sets of Monte Carlo 
     experiments). Only three mutual nodal distance values out of the 1482 computed are below 0.5~au (probability of 0.0020) and 12 are 
     below 1.450~au (probability of 0.0081), the 1st percentile for ${\Delta}_{+}$. Out of these 12 pairs, eight have close ascending nodes 
     and four have close descending nodes (consistent with the asymmetry pointed out above), the two lowest values are for mutual ascending 
     nodes. Out of 39 ETNOs, 41 per cent (16) have at least one mutual nodal distance smaller than 1.450~au. The existence of a subset of 
     pairs with statistically significant small values of the mutual nodal distance suggests that the orbits of some known ETNOs are far 
     from being dynamically uncorrelated. In sharp contrast, there are no pairs with both ${\Delta}_{\pm}$ below 5~au and Fig.~\ref{nodemap} 
     shows that most pairs have (${\Delta}_{+}, {\Delta}_{-}$) around (60~au, 70~au), well beyond the small values of the perhaps 
     dynamically correlated pairs. 
 
     The ETNO pair with the lowest node separation consists of 148209 (2000~CR$_{105}$) and 505478 (2013~UT$_{15}$) with 
     ${\Delta}_{+}=0.23^{+0.25}_{-0.16}$~au (median and 16th and 84th percentiles, the median value is well below the 1st percentile of the 
     distribution) that are part of two different clusters (0 and 2, respectively) as shown in Table~\ref{etnosB}. The second lowest value 
     is found for the pair made of 496315 (2013~GP$_{136}$) and 2015~KG$_{163}$ with ${\Delta}_{+}=0.44^{+0.43}_{-0.31}$~au that again are 
     in different clusters (0 and 1, respectively). The pair consisting of 2002~GB$_{32}$ and 2013~FT$_{28}$ has the third lowest mutual 
     nodal distance with ${\Delta}_{-}=0.48^{+0.36}_{-0.31}$~au, and once more the objects are part of two different clusters (2 and 1, 
     respectively). The pair consisting of 2016~QU$_{89}$ and 2016~SG$_{58}$ has the fourth lowest mutual nodal distance with 
     ${\Delta}_{+}=0.62\pm0.13$~au, although this time the objects are part of the same cluster, \#2. We would like to note that 148209 is 
     part of two relevant pairs, but it is not the only ETNO being part of two or more pairs with low mutual nodal distances, 
     2013~SL$_{102}$ is also part of two and 474640 Alicanto (2004~VN$_{112}$), 505478, and 2016~SG$_{58}$ are part of three. ETNOs 148209 
     and 2016~SG$_{58}$ are part of pairs with both ${\Delta}_{+}$ and ${\Delta}_{-}$ below 1.450~au. Column $X$ in Table~\ref{etnosB} shows 
     the number of pairs with one mutual nodal distance under 1.450~au for each known ETNO. 

     Our geometrical approach indicates that unless a protective mechanism is in place (perhaps in the form of mean-motion or secular 
     resonances but see the comments in Section~1) and considering the uncertainties in the orbit determinations, close encounters at less 
     than 10$^4$~km (at the 5th percentile of the distribution of mutual nodal distances for some pairs, for example 148209 and 505478) may 
     be possible and some of these objects have sizes of hundreds of km so approaches within 1 Hill radius of each other are theoretically 
     achievable. These results open the door to actual collisions at relatively high speeds at perihelion and at aphelion as many known 
     objects have the ascending nodes close to perihelion and the descending ones near aphelion. Whether this may actually be happening or 
     not is difficult to assess because it implies making assumptions on how the Solar system is structured beyond 100~au from the Sun. If 
     there are no perturbers to keep these ETNOs at safe distances from each other, our analysis suggests that collisions may be taking 
     place. Collisional families are not unheard of in the regular trans-Neptunian space, though. The first bona fide asteroid family 
     identified in the outer Solar system was the one associated with dwarf planet Haumea \citep{2007Natur.446..294B} although a candidate 
     collisional family had already been proposed by \citet{2002ApJ...573L..65C} and later confirmed by \citet{2018MNRAS.474..838D} who 
     found four new possible collisional families of TNOs and a number of unbound TNOs that may have a common origin. The subject of 
     finding collisional families of TNOs has been studied by \citet{2003EM&P...92...49C} and \citet{2011ApJ...733...40M}. 

  \section{Poles and perihelia}
\begin{figure}
       \centering
        \includegraphics[width=\linewidth]{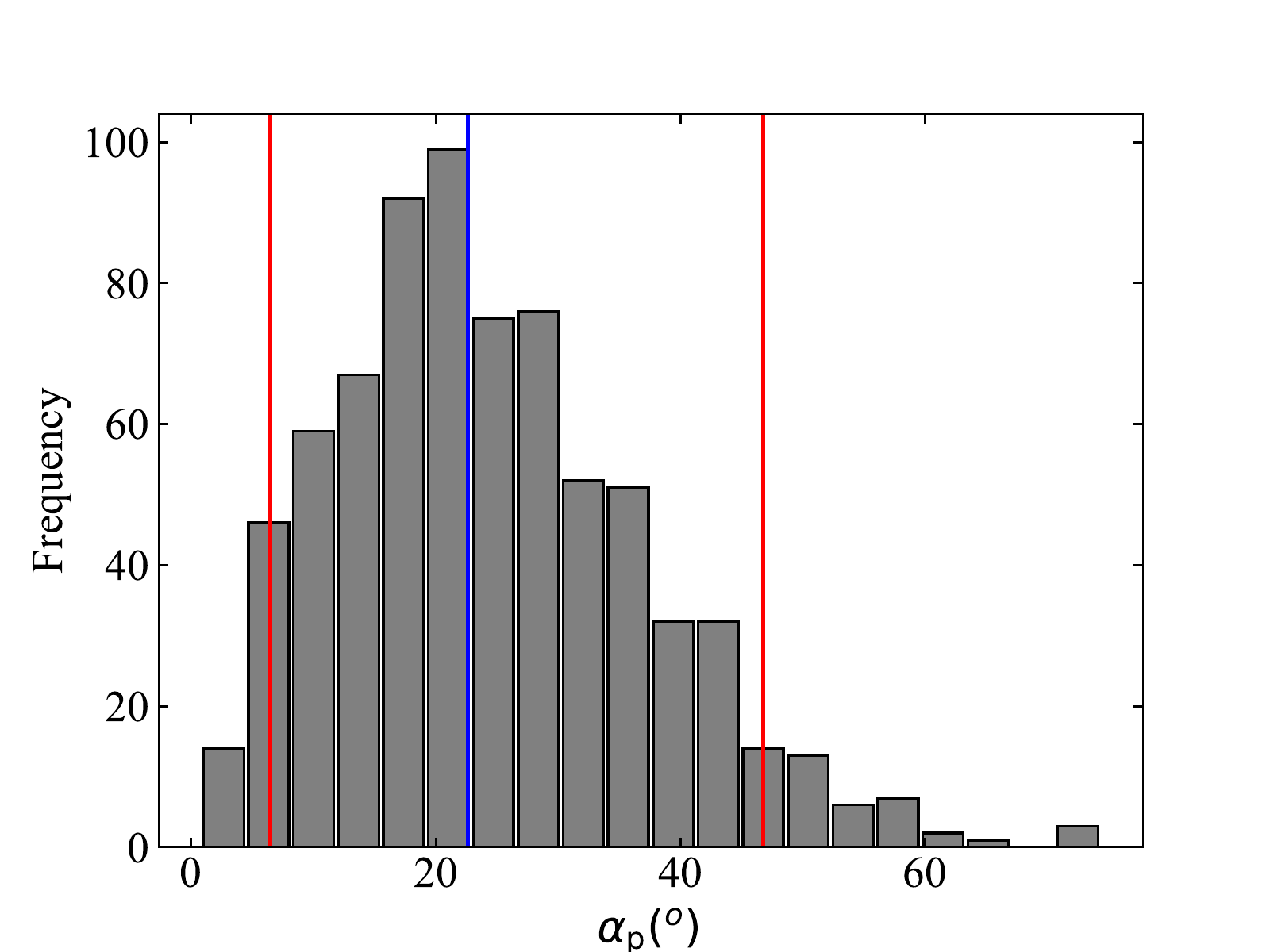}
        \caption{Distribution of angular distances between pairs of orbital poles of the sample of 39 ETNOs. The median is shown in
                 blue and the 5th and 95th percentiles in red. 
                }
        \label{poles}
     \end{figure}

\begin{figure}
       \centering
        \includegraphics[width=\linewidth]{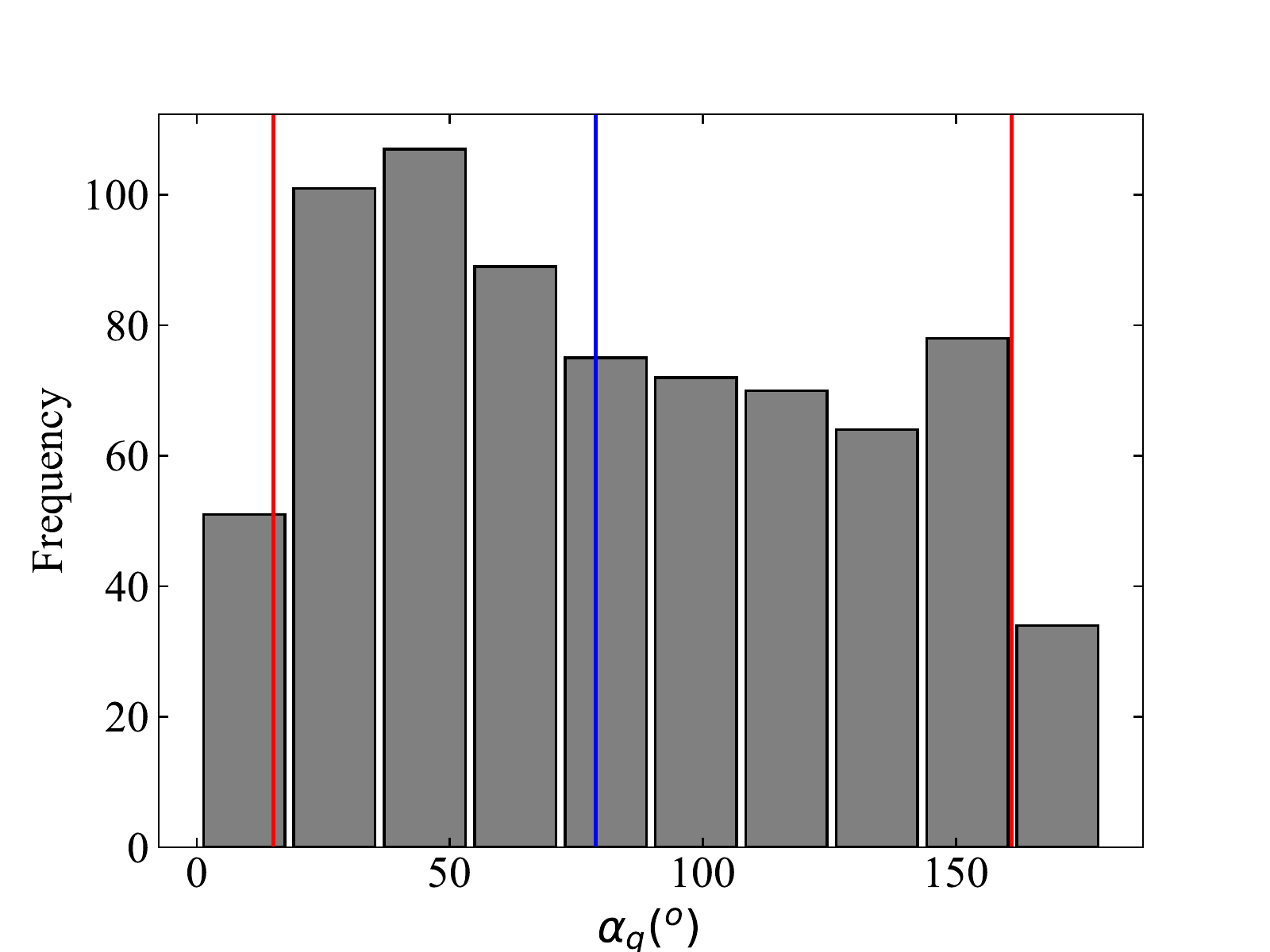}
        \caption{Distribution of angular distances between pairs of perihelia (its projected direction) of the sample of 39 ETNOs. The 
                 median is shown in blue and the 5th and 95th percentiles in red. 
                }
        \label{perihelia}
     \end{figure}

     Here, we focus on the distributions of angular separations between orbital poles, $\alpha_{\rm p}$, and perihelia, $\alpha_q$, of the
     39 ETNOs studied in the previous section. The orbital pole is the intersection between the celestial sphere and a hypothetical axis 
     perpendicular to the plane of the orbit under study. The direction of perihelia is the intersection between the celestial sphere and a 
     hypothetical line that goes from the focus of the orbit towards the point where the orbit under study reaches perihelion. Poles and 
     perihelia are computed as described in Appendix~B. Discarding 2020~KV$_{11}$ again, we have computed 741 pairs of ($\alpha_q$, 
     $\alpha_{\rm p}$) values. The results for each pair come from a set of 10$^4$ pairs of virtual ETNOs as described in Appendix~B. The 
     distribution of angular distances between pairs of orbital poles is shown in Fig.~\ref{poles} and that of the angular distances between 
     pairs of perihelia is displayed in Fig.~\ref{perihelia}. These distributions have been computed using the data in Table~\ref{etnosB} 
     and considering the uncertainties in the orbit determinations as described in Appendixes~A and B.
%
%
%
%
%
%
%
%

\begin{figure}
       \centering
        \includegraphics[width=\linewidth]{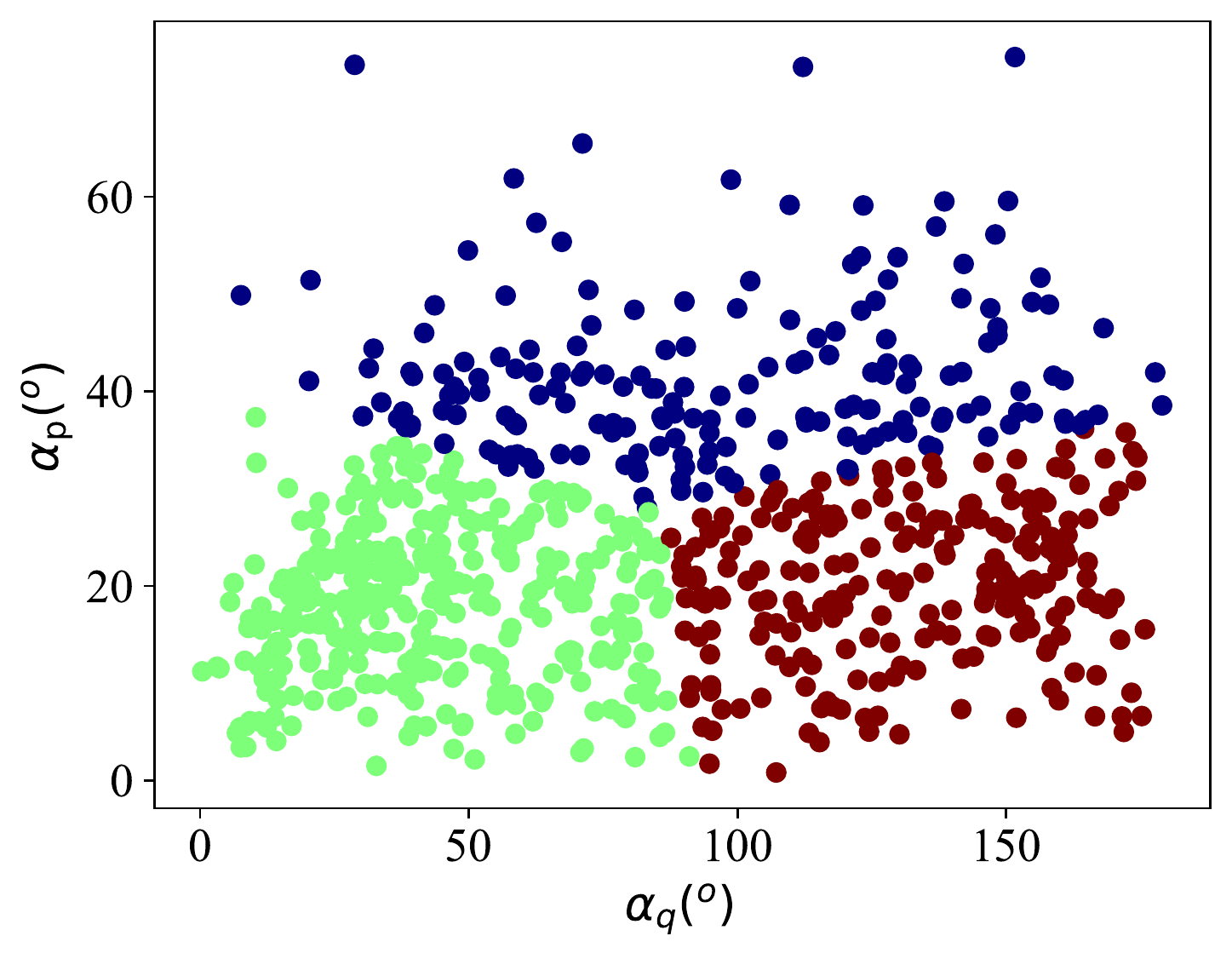}
        \caption{Colour-coded clusters generated by the $k$-means++ algorithm applied to the dataset made of 741 pairs of ($\alpha_q$, 
                 $\alpha_p$) from the sample of 39 ETNOs.
                }
        \label{angclusters}
     \end{figure}

     Applying the $k$-means++ algorithm to the ($\alpha_q$, $\alpha_{\rm p}$) dataset as described in Section~3, we obtained 
     Fig.~\ref{angclusters} that displays three clusters, not four. The values of poles and perihelia provide the orientation of the orbit
     in space and therefore they are directly affected by the observational biases pointed out in Section~1. Nonetheless, the distributions 
     in Figs~\ref{poles} and \ref{perihelia} show that a number of pairs have small values of $\alpha_q$ and $\alpha_{\rm p}$. As we did in 
     the previous section in order to identify severe outliers, we use the 1st percentile of the distribution that is 7\fdg59$\pm$0\fdg02 
     for $\alpha_q$ and 3\fdg260289$\pm$0\fdg000011 for $\alpha_p$ (the 5th percentile values are 15\fdg19 and 6\fdg45, respectively). 
%
%
%
%

\begin{figure}
       \centering
        \includegraphics[width=\linewidth]{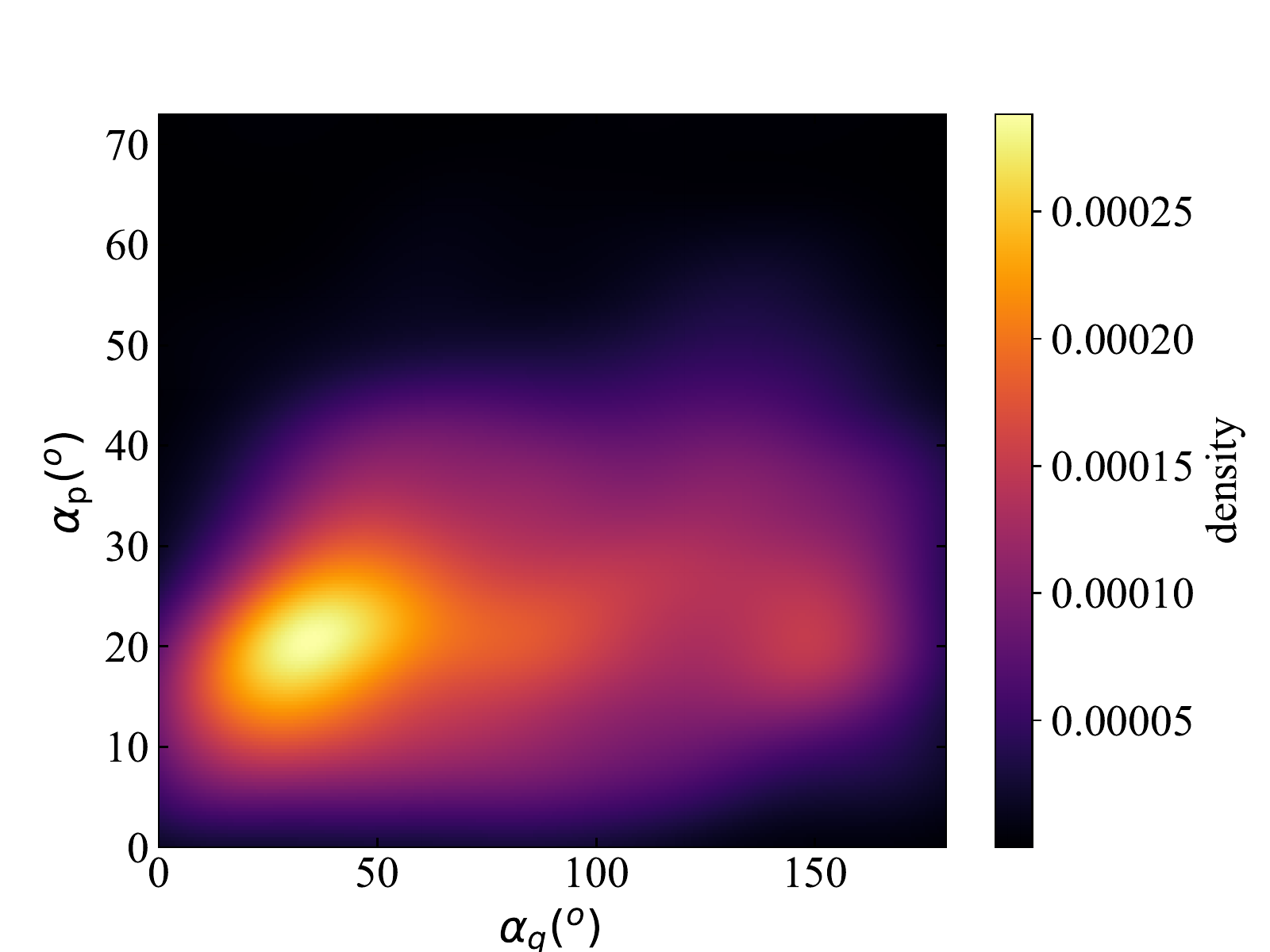}
        \caption{Gaussian kernel density estimation of the same data shown in Figs~\ref{poles} and \ref{perihelia}. 
                }
        \label{ppmap}
     \end{figure}

     There are no pairs with both angular separations under the respective 1st percentile threshold, but there is one pair very close to it: 
     2013~FT$_{28}$ and 2015~KG$_{163}$, both from cluster 1, with $\alpha_q=7\fdg58^{+0\fdg11}_{-0\fdg10}$ and 
     $\alpha_p=3\fdg402\pm0\fdg002$ (median and 16th and 84th percentiles). The pair consisting of 2010~GB$_{174}$ (from cluster 1) and 
     2016~TP$_{120}$ (from cluster 0) has the lowest value of $\alpha_q=0\fdg43^{+0\fdg25}_{-0\fdg10}$ but $\alpha_p=11\fdg219$ and the one 
     including 523622 (2007~TG$_{422}$) and 2016~GA$_{277}$ (from clusters 1 and 2, respectively) has the lowest value of 
     $\alpha_p=0\fdg8260^{+0\fdg0004}_{-0\fdg0003}$ but $\alpha_q=107\fdg2$. In general, the orbits with the closest orientations in space 
     tend to be part of different clusters. Some of the pairs following nearly intersecting orbits as discussed in the previous section also 
     have similar orientations in space. The pair consisting of 148209 (2000~CR$_{105}$) and 505478 (2013~UT$_{15}$) ---from clusters 0 and 
     2, respectively--- that has the lowest node separation also has $\alpha_q=6\fdg26\pm0\fdg02$. Figure~\ref{ppmap} shows that most pairs 
     have ($\alpha_q, \alpha_p$) around (35{\degr}, 20{\degr}), well beyond the small values of the probably correlated pairs.
%
%
%
%

\begin{figure}
       \centering
        \includegraphics[width=\linewidth]{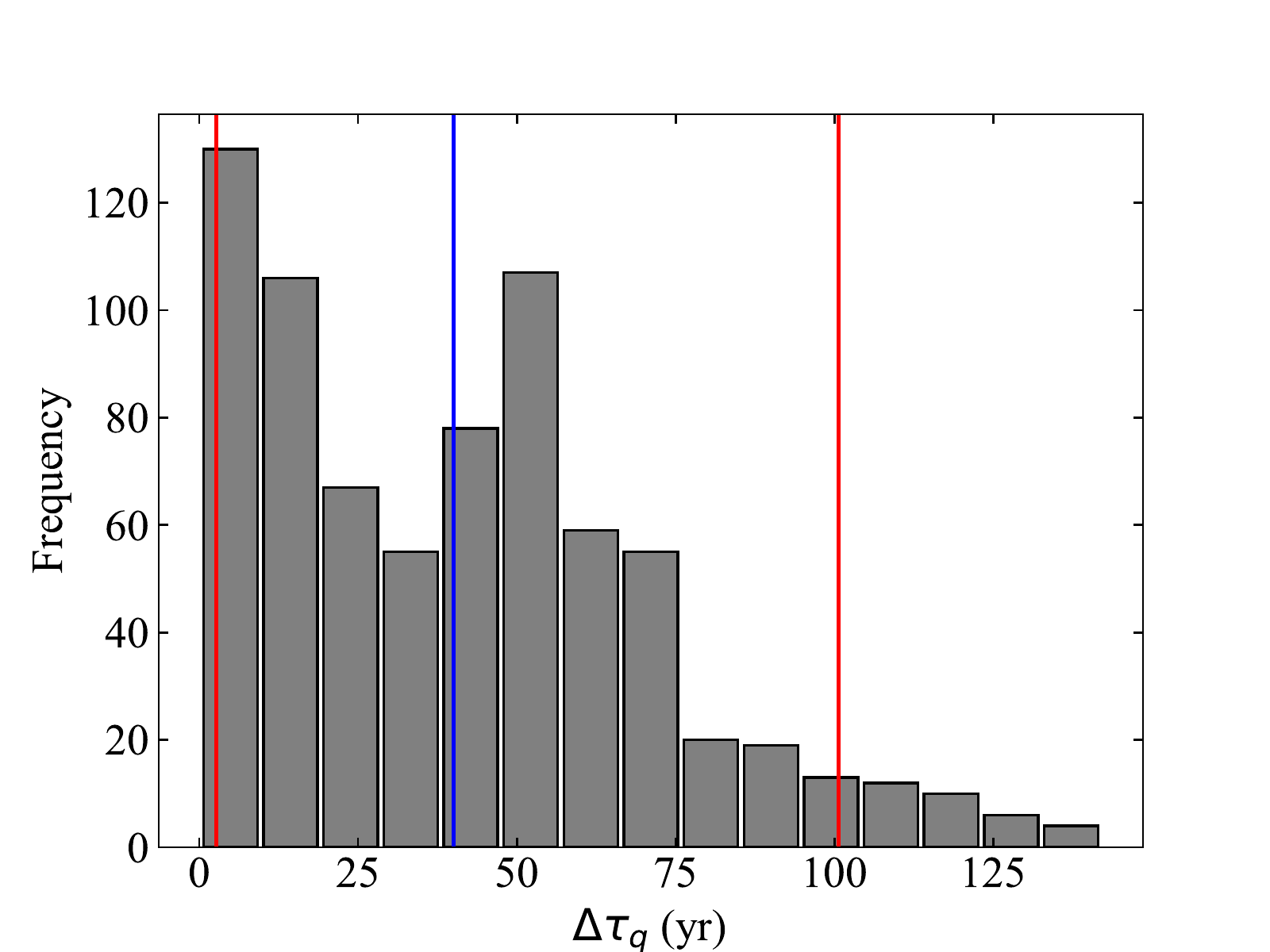}
        \caption{Distribution of differences in time of perihelion passage. The median is shown in blue and the 5th and 95th percentiles in 
                 red.
                }
        \label{timeperpass}
     \end{figure}

     Another quantity of possible interest is the time of perihelion passage, $\tau_q$, not shown in Table~\ref{etnosB} but whose 
     distribution of differences is shown in Fig.~\ref{timeperpass}. In this case, the 5th percentile value is 2.7~yr and the 1st percentile 
     is 0.7~yr. The pair with the lowest value of $\Delta\tau_q$ is the one made of 523622 (2007~TG$_{422}$) of cluster 1 and 527603 
     (2007~VJ$_{305}$) of cluster 2 with $\Delta\tau_q=0.2020\pm0.0005$~yr. The second lowest is for 2013~SY$_{99}$ of cluster 1 and 
     2015~KH$_{163}$ of cluster 0 with $\Delta\tau_q=0.21\pm0.03$~yr. The pair with the third lowest value is the one made of 474640 
     Alicanto (2004~VN$_{112}$) of cluster 0 and 2013~RF$_{98}$ also from cluster 0 with $\Delta\tau_q=0.23_{-0.16}^{+0.23}$~yr. Such low 
     values could be anecdotal as no pairs have very small differences in time of perihelion passage and also very low mutual nodal 
     distances and/or very close angular separations $\alpha_q$ or $\alpha_p$.
%
%
%
%

  \section{Discussion}
     We have performed detailed statistical analyses of a snapshot in orbital parameter space of the known ETNOs, a group of objects whose 
     study may help in understanding how the Solar system is structured beyond 100~au from the Sun. Our analyses do not make any assumptions 
     on how the ETNOs came into existence or how they may be evolving dynamically, they only extract facts associated with this group of 
     objects. It is now time to explore the implications of our findings. 

     The main question initially addressed in this work was that of the underlying level of clustering within the ETNOs. Are there any real 
     clusters or the observed distribution is compatible with an uniform arrangement? This key issue has been previously studied by 
     considering the orbital parameters separately and the results of those analyses have not been conclusive because of the observational 
     bias discussed in Section~1. In order to avoid or at least minimize this problem, we have applied a novel approach by estimating the 
     level of clustering in barycentric ($q_{\rm b}$, $e_{\rm b}$, $i_{\rm b}$) space with the $k$-means++ algorithm. Instead of focusing on 
     one-dimensional clustering, we have applied the algorithm to a three-dimensional dataset. It may have been five-dimensional but we 
     decided to leave the other two angular elements out of the analysis due to the bias concerns discussed in Section~1. The results of 
     this analysis uncover four statistically significant clusters. This is not surprising when considering the case of the regular TNOs 
     that consist of four or more dynamical classes. Therefore and on strictly mathematical grounds, we tentatively consider the answer to 
     the original question as positive: there may be true clustering, the known ETNOs may not represent a single, monolithic population but 
     the mixture of several, probably four, populations. On the other hand, the dynamical classes may be interacting and in this context, we 
     may have to speak of dynamically active and dynamically inert ETNOs (see below, Section~6.1). 

     The actual statistical significance of our results cannot be properly assessed without considering a simplified model for the scattered 
     disc (see below, Sections~6.2 and 6.3). Results from an unbiased model can then be compared with those coming from real ETNOs that are 
     affected by the observational biases pointed out in Section~1. In the following, we have generated a synthetic sample consistent with 
     results obtained by \citet{2001AJ....121.2804B,2017AJ....154...65B} and performed the same analysis carried out in Sections~4 and 5 on 
     the resulting data. This, together with a shuffled version of the actual data serve as reference to estimate the statistical 
     significance of our results. In a shuffled sample and by randomly reassigning the values of the orbital elements of the ETNOs, we 
     preserve the original distributions of the parameters, but remove any possible correlations that may have been induced by hypothetical 
     mean-motion or secular resonances, close encounters with massive perturbers or in response to external perturbations.

     \subsection{Interacting populations: dynamically active versus inert ETNOs}
        In Section~4, we have explored how the members of the candidate clusters interact within their own cluster and with members of other 
        clusters. Two prograde Keplerian trajectories with a common focus can interact only in the vicinity of their mutual nodes, where the
        orbits are closest to each other. We would like to remark that our geometrical approach leaves out of the analysis the current 
        position of the object in its orbit that is controlled by a sixth parameter, the mean anomaly, $M$. We believe that this additional 
        parameter cannot provide significant information in the case of the ETNOs as they are all discovered at perihelion or very near it, 
        so $M$ is always close to 0{\degr} or 360{\degr} for the current sample: the actual intervals are (0\fdg39, 12\fdg44) and 
        (348\fdg07, 359\fdg97). The analysis presented in Section~4 shows that the nodal distances organize themselves into four clusters 
        that we consider consistent with the four clusters found in barycentric ($q_{\rm b}$, $e_{\rm b}$, $i_{\rm b}$) space. 

        The analysis also uncovers that out of 39 ETNOs, 16 or 41 per cent have at least one mutual nodal distance smaller than 1.45~au that 
        is the 1st percentile of the distribution. We call these objects dynamically active ETNOs and they are identified in 
        Table~\ref{etnosB} by a value of $X>0$, the number of pairs with at least one mutual nodal distance smaller than 1.45~au in which 
        they are members; dynamically inert ETNOs have $X=0$ in Table~\ref{etnosB}. We believe that this result is quite remarkable when 
        considering the large size of the orbits studied here. It is difficult to attribute this finding to mere coincidence and some 
        processes should be operating in the background to lead to median values of the mutual nodal separation under 0.5~au in pairs such 
        as 148209 (2000~CR$_{105}$) and 505478 (2013~UT$_{15}$). The fraction of dynamically active ETNOs considering the simple scattered 
        disc model discussed in the next section is 54 per cent and the one from an experiment with shuffled orbital parameters is also 
        54 per cent. This result, obtained randomly sampling 39 orbits 10$^{4}$ times, strongly suggests that, in this regard, shuffling the 
        observed orbital parameters is equivalent to considering the simple scattered disc model. 

        In addition, some pairs of orbits (for example, those of 148209 and 505478) have non-negligible probabilities (small but $>5$ per 
        cent) of approaching each other under a few thousand kilometers when uncertainties are factored in. This fact can be signalling the 
        existence of collisions or of mechanisms to avoid them such as mean-motion or secular resonances. If we study the dynamically active 
        fraction in the samples of 10$^{4}$ synthetic ETNOs (so 49995000 pairs), we find that 100 per cent of the objects have at least one 
        mutual nodal distance smaller than 1.45~au; in fact, each object is part of over 200 pairs in which one nodal distance is smaller 
        than 1.45~au. This result together with the previous one of an active fraction of 54 per cent when samples of 39 objects are 
        considered suggests that there is a probable excess of dynamically inert ETNOs. 

        Within the context of the simple scattered disc model, physical collisions or orbit-changing close encounters (if massive ETNOs are 
        involved) could be ubiquitous unless protective mechanisms are in place. Close encounters and collisions in the trans-Neptunian or 
        Kuiper belt are relatively well understood (see e.g. \citealt{2013A&A...558A..95D,2020AJ....160...85B,2020PSJ.....1...40K,
        2021AJ....161..195A}). Collisional families are known to populate regular TNO space that is also affected by orbital resonances with 
        small integer ratios and secular resonances like the so-called von~Zeipel--Lidov--Kozai oscillation \citep{1910AN....183..345V,
        1962P&SS....9..719L,1962AJ.....67..591K,2019MEEP....7....1I}. Resonances are driven by massive perturbers that, in the case of 
        regular TNOs, have Neptune as main driver and the other giant planets as secondary perturbers. Close encounters with a massive 
        perturber can also produce pairs that resemble those observed via binary stripping \citep{2017Ap&SS.362..198D}. 
        
        No massive perturbers are known to exist in ETNO space, but the exploration of this vast region is just beginning (see e.g. 
        \citealt{2018AJ....156..135H,2018A&A...615A.159P,2019AJ....158...43K,2019EPSC...13.1528S,2020ApJS..247...32B,2020EPSC...14..420R,
        2021arXiv210410264N}). \citet{2014Natur.507..471T} argued that the orbital organization of the few ETNOs known in 2014 was 
        indicative of the presence of one hidden massive perturber at hundreds of au from the Sun. However, \citet{2014MNRAS.443L..59D} and 
        \citet{2015MNRAS.446.1867D} contended that the available data at the time were better explained if more than one massive perturber 
        was orbiting the Sun well beyond Pluto. The size of the ETNO dataset was still limited to less than a dozen objects when 
        \citet{2016AJ....151...22B} presented the so-called Planet~9 hypothesis as an explanation for the orbital architecture of the known 
        ETNOs. The discovery of additional objects led to modifications of the original hypothesis. 

        The latest incarnation of the Planet~9 hypothesis \citep{2019PhR...805....1B} predicts the existence of a planet with a mass in the 
        range 5--10~$M_{\oplus}$, following an orbit with a value of the semimajor axis in the range of 400--800~au, eccentricity in the 
        range of 0.2--0.5, and inclination in the interval between (15{\degr}, 25{\degr}). However, \citet{2020A&A...640A...6F} used the 
        INPOP19a planetary ephemerides that include Jupiter-updated positions by the Juno mission and a reanalysis of Cassini observations 
        to conclude that if Planet~9 exists, it cannot be closer than 500~au, if it has a mass of 5~$M_{\oplus}$, and no closer than 650~au, 
        if it has a mass of 10~$M_{\oplus}$.  A recent study using Monte Carlo random search techniques \citep{2021A&A...646L..14D} has 
        found that if the high eccentricities of the known ETNOs are the result of relatively recent flybys with putative planets, these 
        perturbers might not be located farther than 600~au and they have to follow moderately eccentric and inclined orbits to be capable 
        of experiencing close encounters with multiple known ETNOs. This result places the perturbers in the so-called inert Oort Cloud 
        proposed by \citet{2019A&A...629A..95S}. Planetary bodies may have been inserted there early in the history of the Solar system via 
        planetary scattering (see e.g. \citealt{2016ApJ...826...64B,2019AJ....158...94B}) or, less likely, formed in situ (see e.g. 
        \citealt{2015ApJ...806...42K,2016ApJ...825...33K}).

        Focusing on how the ETNOs with one mutual nodal distance smaller than 1.45~au are distributed across the clusters in 
        Table~\ref{etnosB} (those with $X>0$), there are six from cluster 0 (6/11, 54.5 per cent), five from cluster 1 (5/12, 41.7 per 
        cent), five from cluster 2 (5/15, 33.3 per cent), and none from cluster 3, Sednoids (0/2, 0 per cent). Six pairs include a member of 
        cluster 1, ten pairs include members of cluster 0, nine pairs include members of cluster 2, and there are no pairs including members 
        of cluster 3. Within each cluster there are dynamically inert members whose orbits appear to be well detached from all the other 
        orbits and dynamically active members that seem to have the theoretical capability of interacting with members of the same cluster 
        (seldom) and of other clusters (more often) as seen in Figs ~\ref{ascnodesALL} and \ref{desnodesALL}. These dynamically active 
        members include 148209, 474640 Alicanto (2004~VN$_{112}$), 505478, 2013~SL$_{102}$ and 2016~SG$_{58}$. The cluster with the largest 
        number of dynamically active members is 0 and the one with the smallest is 3; the smallest cluster, the Sednoids, does not have any 
        dynamically active members. 

        The case of Alicanto and 2013~SL$_{102}$ is particularly indicative of the existence of some level of gravitational interaction 
        because they have a short mutual ascending nodal distance, ${\Delta}_{+}=0.66^{+0.37}_{-0.36}$~au, and a small difference in time of 
        perihelion passage, $\Delta\tau_q=0.71\pm0.03$~yr, therefore they may pass relatively close to each other within a relatively narrow 
        time window as their orbits have nearly synchronized perihelion passages. As pointed out above, the 1st percentile of the 
        distribution of differences in time of perihelion passage is 0.7~yr so the value of $\Delta\tau_q$ for this pair is barely beyond 
        the outlier boundary considered in this work. In sharp contrast, objects like Sedna could be considered as dynamically inert. It is 
        unclear why do we have two distinct groups regarding mutual nodal distances, but this is unlikely to be accidental or due to 
        observational biases. The cluster with the smallest number of dynamically active members is cluster 3, it is also the one with the 
        least members. The properties of cluster 1 make it compatible with an origin in the so-called inner Oort Cloud 
        \citep{2001AJ....121.2253L} that was first proposed by \citet{1981AJ.....86.1730H}. 

        When considering the dynamically active objects, if they are somehow related, it is unlikely that they formed during a gentle split 
        (i.e. the relative speed at separation could have been low, slightly above the escape velocity from the surface of the most massive 
        object) given the range of values of the various parameters discussed in the previous sections. A more plausible scenario may be 
        that of a high speed collision with catastrophic effects producing two clouds of sizeable fragments moving along the original paths 
        that eventually spread throughout the colliding orbits. In this scenario, two large fragments undergoing a flyby at a later time may 
        have different surface properties and their relative velocity at the distance of closest approach could be close to the original 
        impact speed as they come from the two original, likely unrelated impactors (for example, consider the case of 148209 and 505478). 
        There are few spectroscopic results published on the surface composition of the ETNOs. The results in \citet{2017MNRAS.467L..66D} 
        show that the spectral slopes of Alicanto and 2013~RF$_{98}$, both in cluster 0 and with one of the lowest values of $\Delta\tau_q$, 
        are very close matches and compatible with the ones of 2002~GB$_{32}$ and 506479 (2003~HB$_{57}$), both in cluster 2. However, they 
        are very different from that of Sedna that is the reference object in cluster 3. The spectral slope could in principle be used to 
        separate groups, but if the orbital evolution in ETNO space is chaotic due to the presence of some yet-undetected perturbers (or any 
        other capable dynamical mechanism), true dynamical classes may host objects with different spectral slopes. Considering together 
        spectral and orbital information may help in understanding better how the various objects could be related, but published spectral 
        data on the ETNOs are still scarce.

     \subsection{Statistical significance}
\begin{figure}
       \centering
        \includegraphics[width=\linewidth]{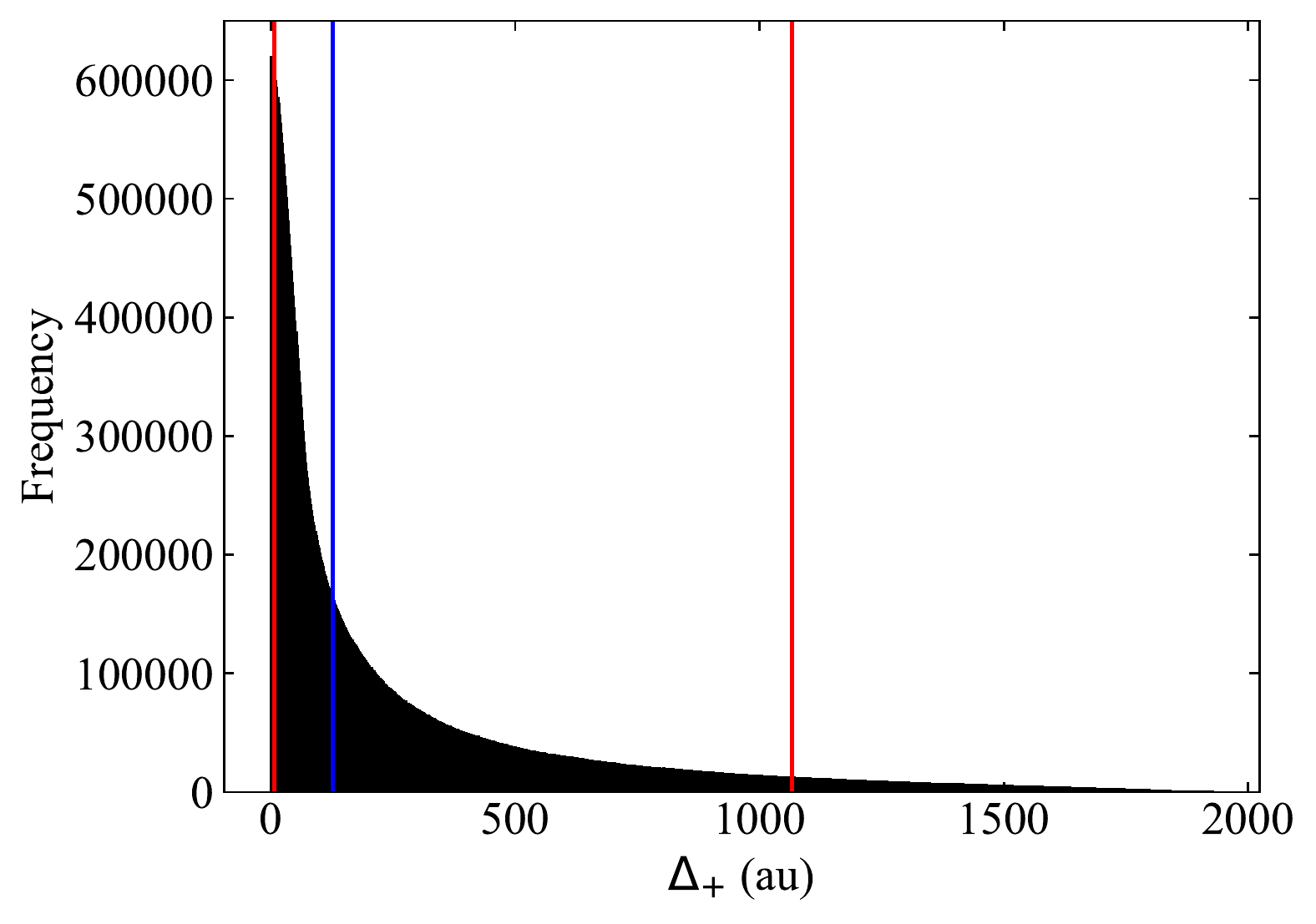}
        \includegraphics[width=\linewidth]{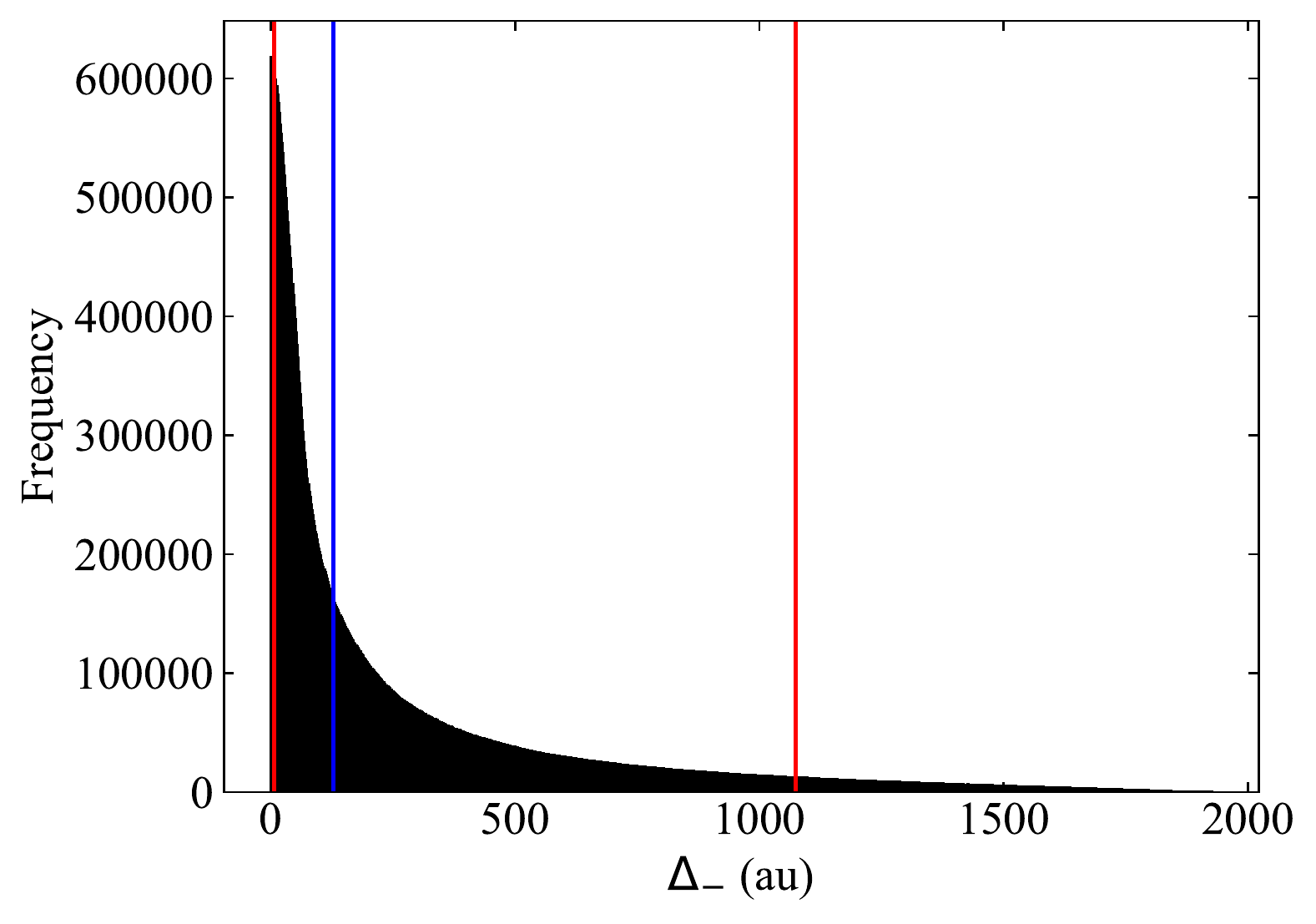}
        \caption{Distribution of mutual nodal distances of ascending nodes (top panel) and of descending nodes (bottom panel) of synthetic 
                 ETNOs. The median is shown in blue and the 5th and 95th percentiles in red. 
                }
        \label{syntnodes}
     \end{figure}

        Let us consider a scattered disc model in which $a$ follows a uniform distribution in the interval (150, 1000)~au, $q$ is also 
        uniform in the interval (30, 100)~au, the angular elements $\Omega$ and $\omega$ are drawn from a uniform distribution in the 
        interval (0\degr, 360\degr), and $i$ follows the so-called Brown distribution of inclinations \citep{2001AJ....121.2804B,
        2017AJ....154...65B} in the interval (0\degr, 60\degr) as described in Appendix~C. After analyzing, as pointed out above, 10 
        instances of 10$^{4}$ synthetic ETNOs each (or 49995000 pairs per instance), we found that the 1st percentile of ${\Delta}_{+}$ is 
        1.460$\pm$0.009~au and the one for ${\Delta}_{-}$ is 1.450$\pm$0.009~au. Therefore and under the assumption of a uniform scattered 
        disc model with a Brown distribution of inclinations, we expect a symmetrical distribution of mutual nodal distances; in other 
        words, the median and percentiles of the mutual ascending and descending nodal distances must be identical (see 
        Fig.~\ref{syntnodes}). 
%
%
%
%

        On the other hand, we observe that the values of the 1st percentile of ${\Delta}_{+}$ for the simple model and the one of the actual 
        data are virtually identical (1.1$\sigma$ difference, considering the model as reference). In sharp contrast and for ${\Delta}_{-}$, 
        we find a difference of over 98$\sigma$ between the model and the real data. It is difficult to conclude that such a consistent 
        match in ${\Delta}_{+}$ could be the result of chance alone; the simple scattered disc model discussed above must, in some way, 
        resemble the actual distributions of orbital elements of the ETNOs.  

\begin{figure}
       \centering
        \includegraphics[width=\linewidth]{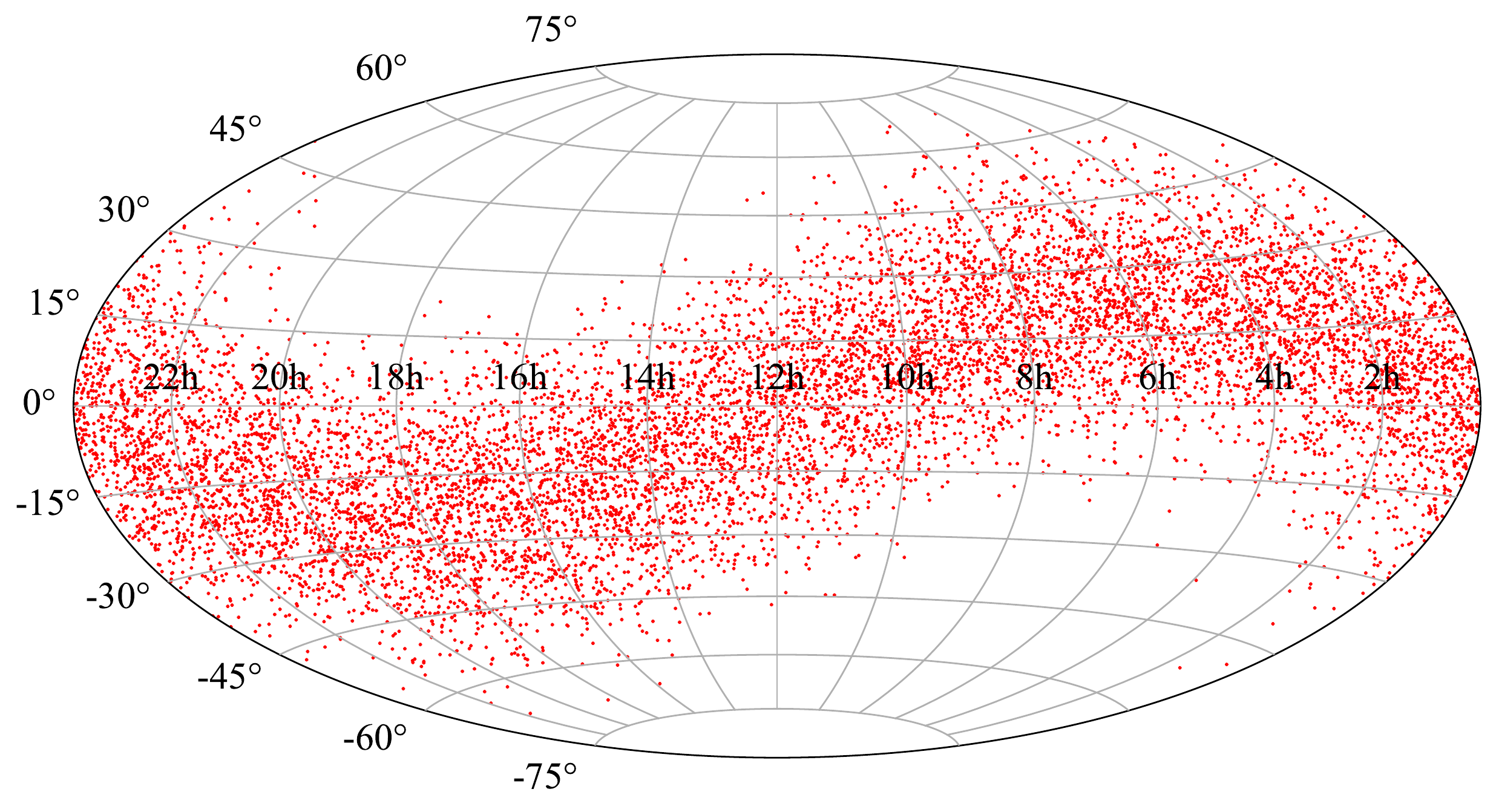}
        \caption{Distribution of the missing synthetic ETNOs that may have produced pairs with ${\Delta}_{-}$ under 2.335~au according to 
                 the model discussed in Section~6.1.
                }
        \label{missing}
     \end{figure}

        As for the strong asymmetry (model versus real data) found between the match in the 1st percentile of ${\Delta}_{+}$ and the large
        difference in the 1st percentile of ${\Delta}_{-}$, the easiest explanation could be that it may be linked to the biases in the 
        angular elements discussed above. This, however, has strong implications: the missing low values of the mutual descending nodal 
        distances must correspond to pairs that have not been found by the various surveys contributing discoveries of this type of objects. 
        Figure~\ref{missing} shows the predicted locations at perihelia (where the known ETNOs have been found) of such missing objects 
        (the ones with ${\Delta}_{-}<2.335$~au) according to the model. They outline the ecliptic but in a rather random fashion so it is 
        difficult to assume that the statistically significant asymmetry between the shortest mutual ascending and descending nodal 
        distances present in the data could be the exclusive result of observational bias, other effects must be at work. In addition, the
        asymmetry persists when considering 10$^{4}$ samples of 39 orbits obtained by shuffling the orbital parameters in Table~\ref{etnosB}. 
        The observational biases discussed in the literature (see Section~1) affect the angular elements but not $a$ or $e$ and, therefore, 
        not $q$. The asymmetry might have arisen in response to external perturbations.
%
%
%
%

\begin{figure}
       \centering
        \includegraphics[width=\linewidth]{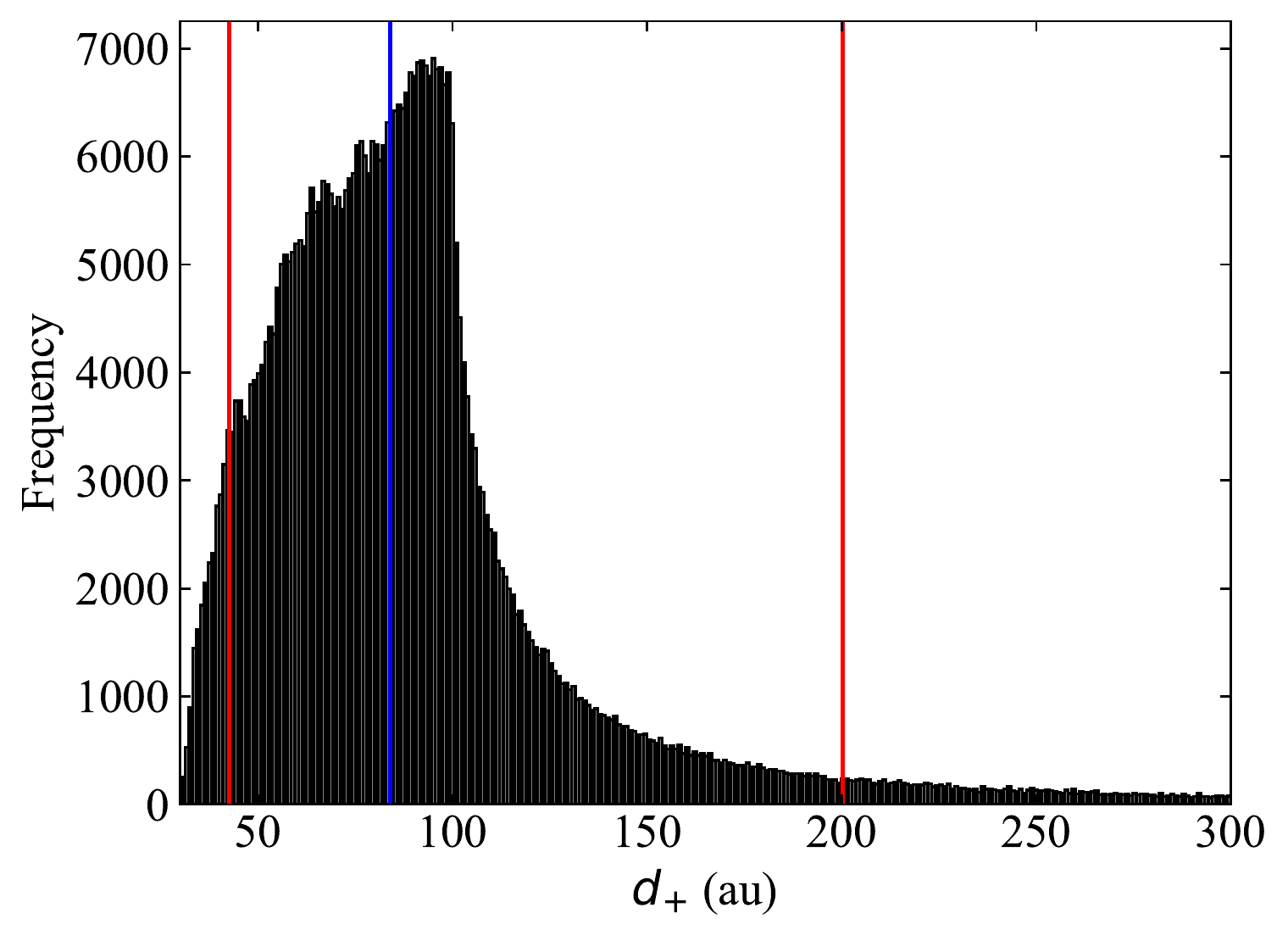}
        \includegraphics[width=\linewidth]{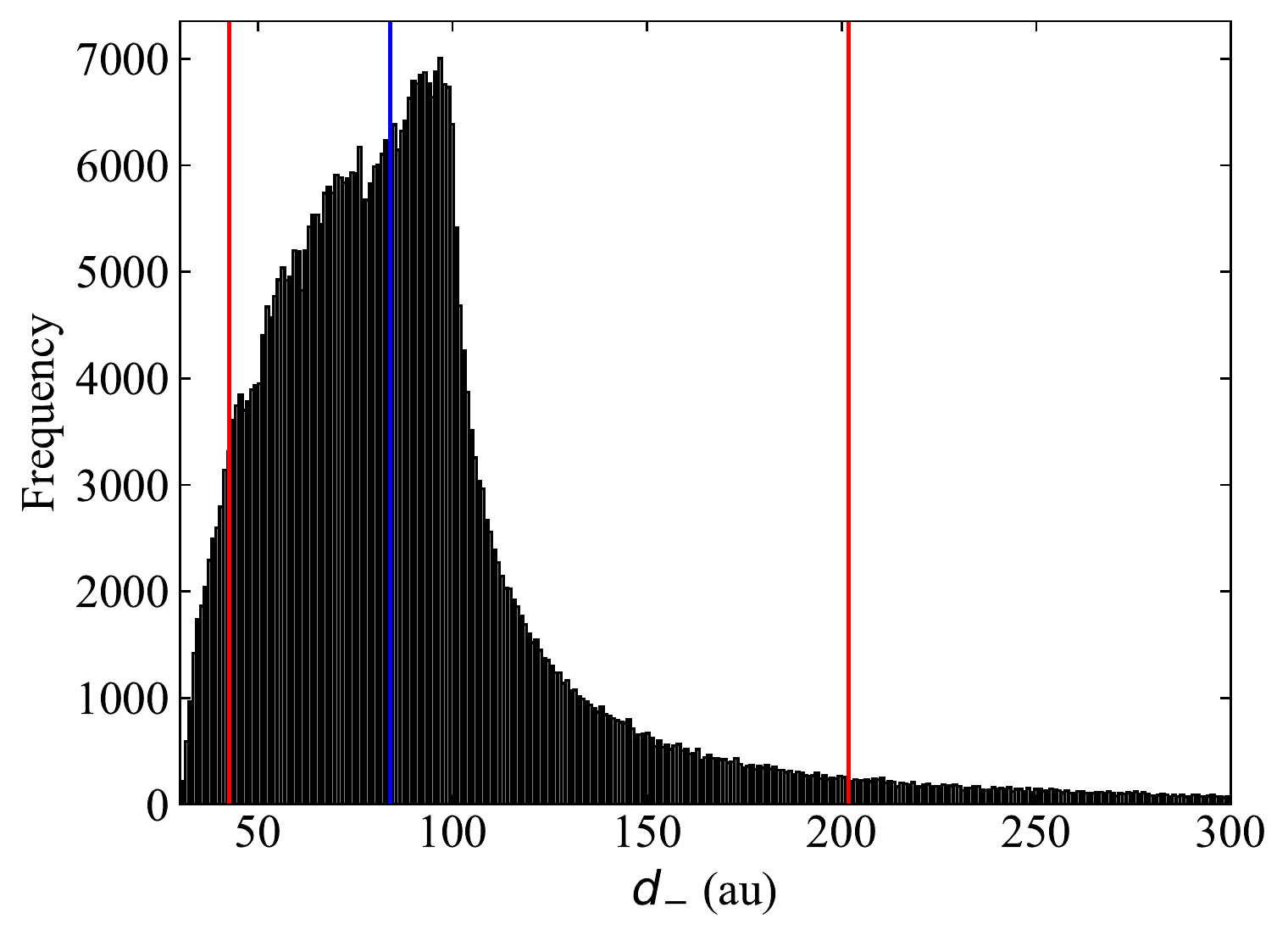}
        \caption{Distribution of barycentric distances to mutual ascending nodes (top panel) and descending nodes (bottom panel) of synthetic 
                 ETNOs (10$^{4}$). The median is shown in blue and the 5th and 95th percentiles in red. 
                }
        \label{syntdnodes}
     \end{figure}

        Continuing with the predictions from the simple scattered disc model discussed above, the pair of ETNOs in cluster 2 made of 505478 
        (2013~UT$_{15}$) and 2016~SG$_{58}$ has ${\Delta}_{+}=1.35_{-0.95}^{+1.37}$~au at 339~au from the barycentre of the Solar system.
        The probability of finding a pair of synthetic ETNOs with ${\Delta}_{+}<1.450$~au beyond 300~au is 0.000205$\pm$0.000005 (the odds 
        are nearly 1 in 5000 so one has to have 5000 pairs, not 741, to be able to find such a distant close nodes). Therefore, if the 
        observed ETNOs come from a distribution similar to the simple scattered disc model used in this section, this pair is truly unusual 
        and should not have been observed in a sample of just 39 objects (see the distribution of distances to the nodes in 
        Fig.~\ref{syntdnodes}). If we randomly sample 39 orbits 10$^{4}$ times from the simple scattered disc model and by shuffling the 
        orbital parameters, we obtain a consistent result: that no pairs with close mutual nodes must exist beyond 300~au and that the 99th 
        percentile of the distribution of barycentric nodal distances is $\sim$110~au.
%
%
%
%

\begin{figure}
       \centering
        \includegraphics[width=\linewidth]{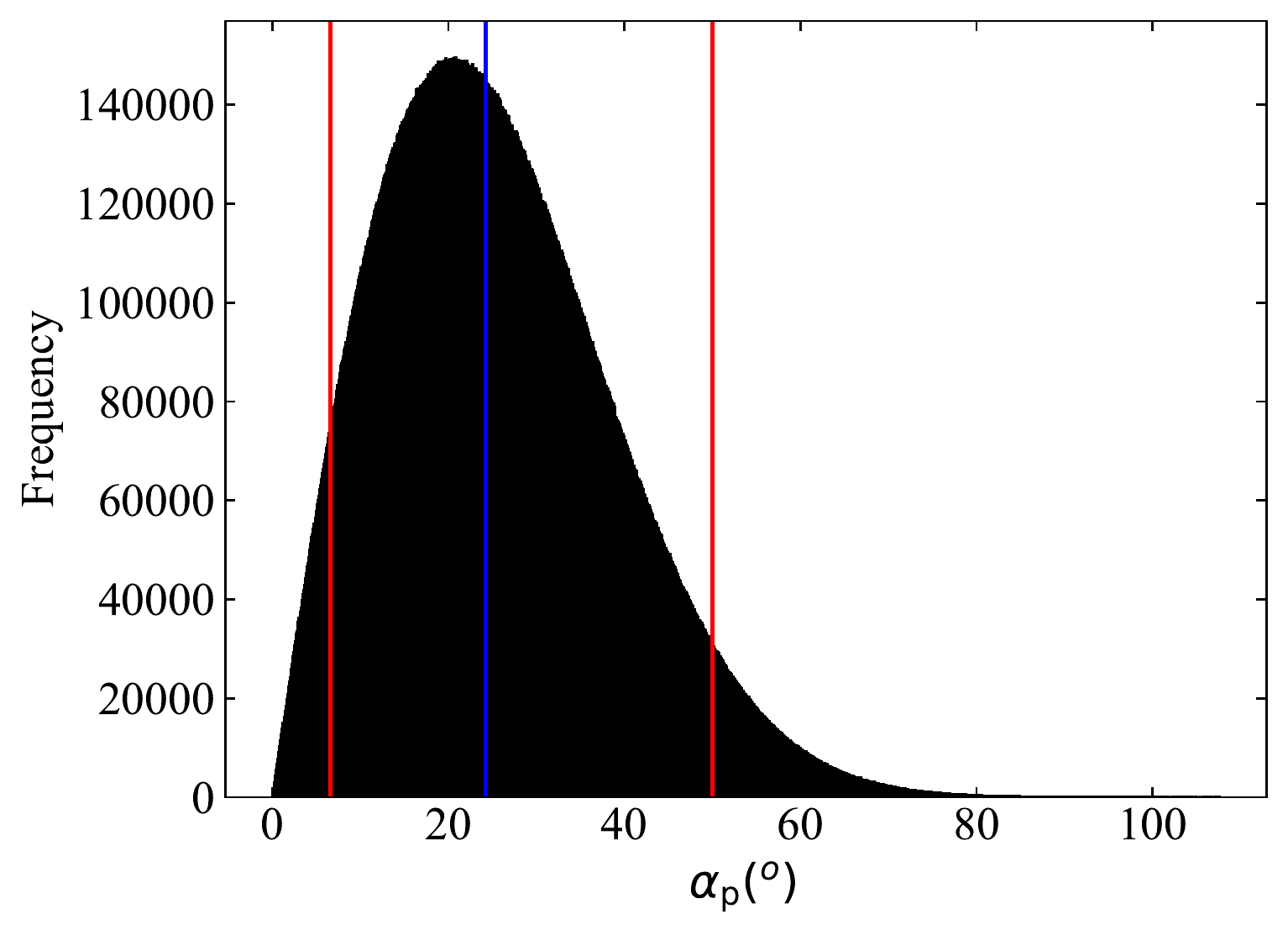}
        \includegraphics[width=\linewidth]{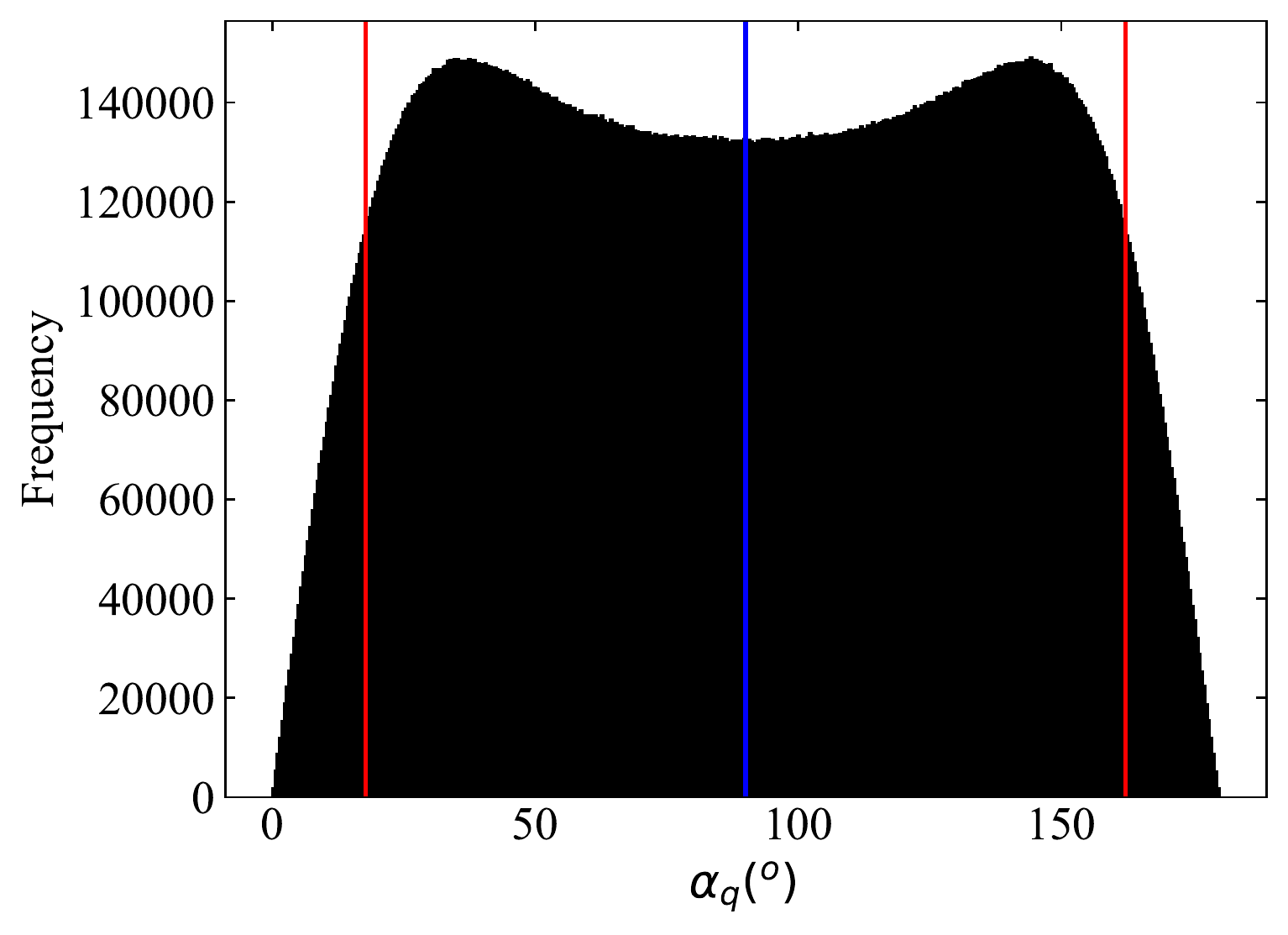}
        \caption{Distribution of angular distances between pairs of orbital poles (top panel) and of angular distances between pairs of 
                 perihelia (its projected direction, bottom panel) of synthetic ETNOs (10$^{4}$). The median is shown in blue and the 5th 
                 and 95th percentiles in red. 
                }
        \label{syntori}
     \end{figure}

        As for the orientations in space of the orbits, the 1st percentile of $\alpha_q$ from the model is 7\fdg70$\pm$0\fdg02 and
        2\fdg9$\pm$0\fdg2 for $\alpha_p$. In this case, the discrepancies between model an real data in the low-probability domain are 
        less pronounced (5.5$\sigma$ and 1.8$\sigma$, respectively). There is however an obvious issue when comparing the observational 
        distribution in $\alpha_q$ (see Figs~\ref{poles}--\ref{ppmap}) and that of the synthetic ETNOs (see Fig.~\ref{syntori}, bottom 
        panel): the second maximum at about 145{\degr} is missing from the distribution. As the distributions of $\alpha_q$ and $\alpha_p$ 
        only depend on the angular elements and these are affected by observational biases, the most plausible explanation for the missing 
        maximum is in the biases themselves.         
%
%
%
%

        In order to check for internal consistency, we have performed the analysis of 10$^6$ samples of 39 ETNOs obtained by randomly 
        scrambling or shuffling the orbital parameters in Table~\ref{etnosB} and another set of 10$^6$ samples of 39 synthetic ETNOs 
        obtained as pointed out in this section. The results from both sets are virtually the same and we interpret this outcome as 
        supportive of two conclusions: that the simple scattered disc model could be good enough as reference and that the known ETNOs 
        are affected by hidden orbital correlations that may not be the exclusive result of observational biases. 

     \subsection{Caveats}
\begin{figure}
       \centering
        \includegraphics[width=\linewidth]{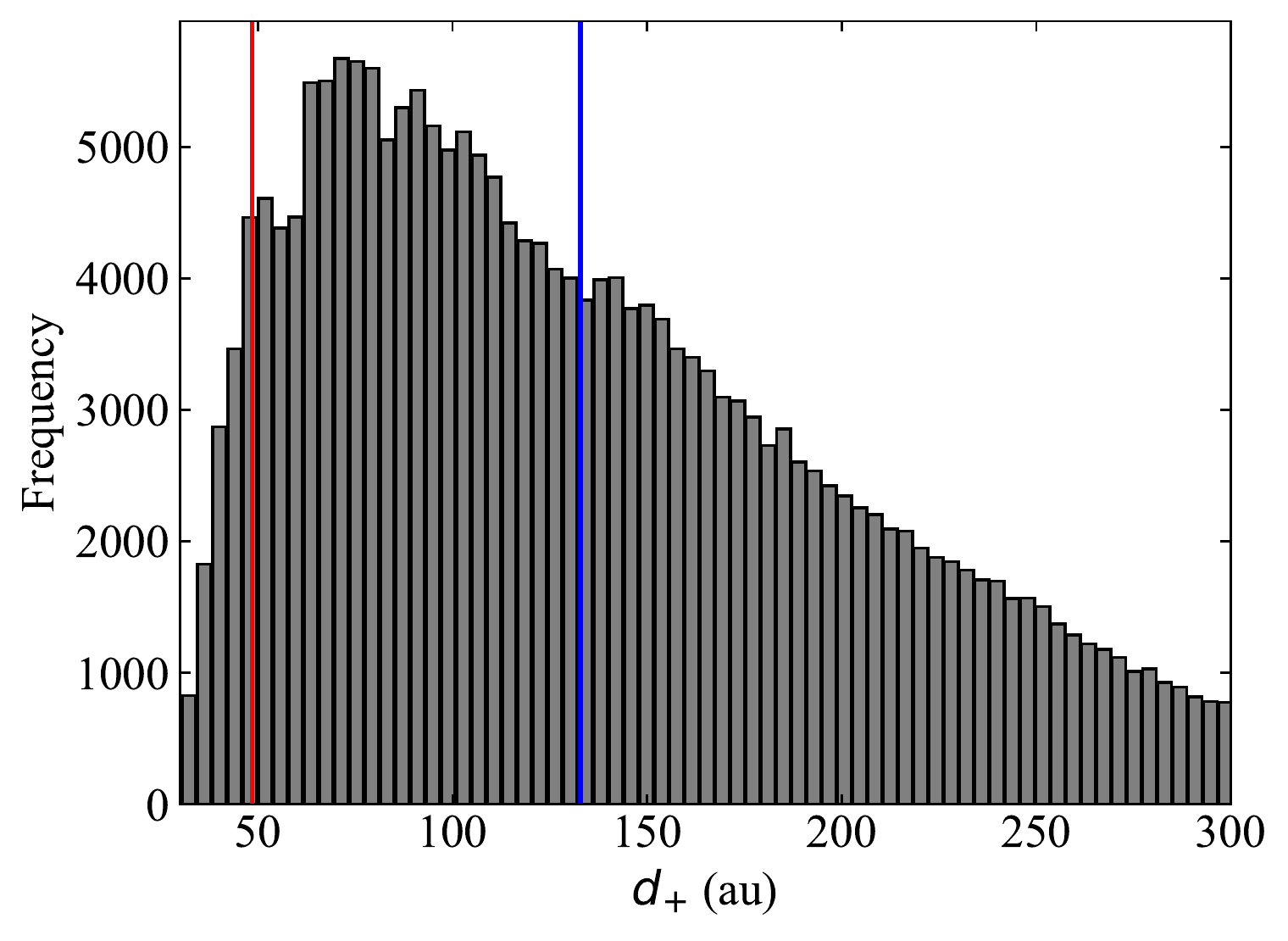}
        \includegraphics[width=\linewidth]{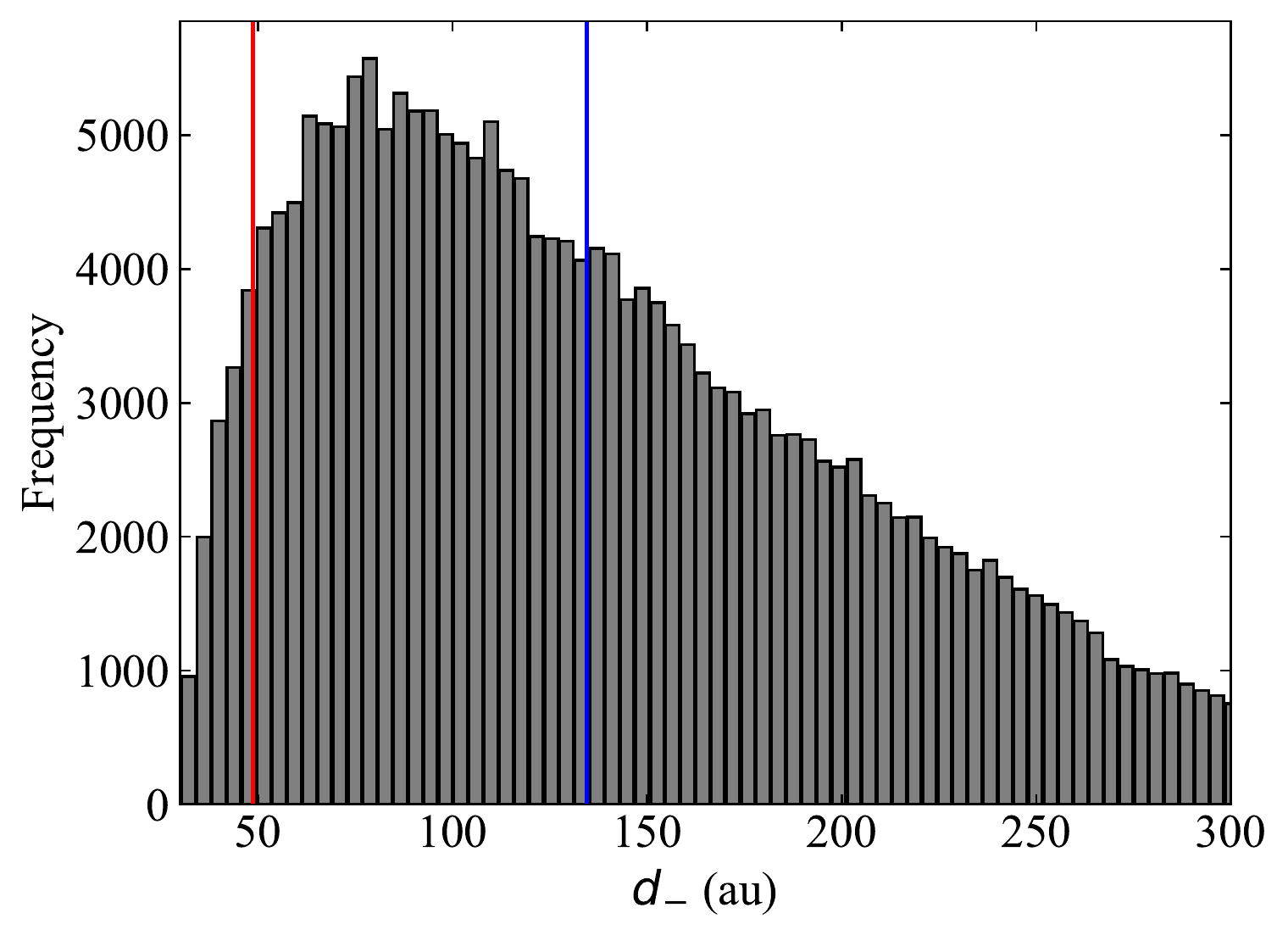}
        \caption{As Fig.~\ref{syntdnodes} but for the model in \citet{2021PSJ.....2...59N}. 
                }
        \label{syntdnodesBN}
     \end{figure}

        There are a number of caveats which concern the robustness of the statistical significance estimates computed in the previous 
        section. The simple scattered disc model discussed above is just one of many that may be used as a reference. 
        \citet{2021PSJ.....2...59N} argue for a model in which $a$ follows the distribution $N(a) \propto a^{0.7}$, $e$ is uniformly 
        distributed in the interval (0.69, 0.999), $i$ follows the Brown distribution as before, and the angular elements $\Omega$ and 
        $\omega$ are also drawn from a uniform distribution in the interval (0\degr, 360\degr). Under such model, the orientations in space 
        of the orbits ($\alpha_q$, $\alpha_p$), remain unchanged, but the distributions of ${\Delta}_{\pm}$ are quite different as the 
        disc is significantly less concentrated than in the case discussed in the previous section (see Fig.~\ref{syntdnodesBN}). For this
        model (also 10$^{4}$ synthetic ETNOs), the 1st percentile of ${\Delta}_{+}$ is 2.98~au and the one for ${\Delta}_{-}$ is 2.97~au. 
        The asymmetry pointed out above remains and now the unusually low values of ${\Delta}_{\pm}$ presented above are even more 
        significant (less probable). The probability of finding a pair of synthetic ETNOs in this model with ${\Delta}_{+}<1.450$~au beyond 
        300~au is 0.00047$\pm$0.00001 that is still very low, although higher than the one obtained with the model discussed in the 
        previous section.
%
%
%
%

        The origin of the improbable orbits (in terms of correlations) observed and the statistically robust asymmetry present in the 
        distribution of mutual nodal distances computed for the known ETNOs is open to discussion. Collisions and the effects of 
        yet-to-be-discovered massive perturbers (see above) may have a role in explaining what is observed as well as stellar encounters and 
        (less likely) the Galactic tide (see e.g. \citealt{2018ApJ...863...45P}). The warped midplane of the trans-Neptunian belt discussed 
        by \citet{2017AJ....154...62V,2017AJ....154..212V} might also evidence past stellar encounters (see e.g. 
        \citealt{2018MNRAS.476L...1D,2020ApJ...901...92M}) and/or be linked to the asymmetry uncovered by our analyses.   

  \section{Conclusions}
     In this paper, we have applied machine-learning techniques to explore the possible existence of clusterings or subgroups within the 
     ETNO group of asteroids (those with $a>150$~au and $q>30$~au). Then, we have investigated the presence of correlations using the 
     distributions of the nodal distances and the poles and perihelia of the orbits within the context of the uncertainties of their 
     barycentric orbit determinations. The main conclusions of our study are:
     \begin{enumerate}[(i)]
        \item We identify four statistically significant groups or clusters in ($q_{\rm b}$, $e_{\rm b}$, $i_{\rm b}$) ETNO orbital 
              parameter space by applying the $k$-means++ algorithm. 
        \item One of the statistically significant groups includes 90377 Sedna (2003~VB$_{12}$) and 2012~VP$_{113}$ and appears to be well
              dynamically detached from the rest of the ETNOs.
        \item Another statistically significant group may be related to the inner Oort cloud and includes objects like 541132 Leleakuhonua 
              (2015~TG$_{387}$), 2013~SY$_{99}$, 2014~FE$_{72}$, and 2015~KG$_{163}$. 
        \item The distribution of mutual nodal distances shows a statistically significant asymmetry between the shortest mutual ascending 
              and descending nodal distances. We argue that this statistically robust asymmetry is not due to observational bias but perhaps 
              the result of external perturbations.
        \item Although most nodal distances are found in the range 60--70~au, out of 39 ETNOs, 16 or 41 per cent have at least one mutual 
              nodal distance smaller than 1.45~au that is the 1st percentile of the distribution. There are no pairs with both mutual nodal 
              distances smaller than 5~au.
        \item The peculiar pair of ETNOs made of 505478 (2013~UT$_{15}$) and 2016~SG$_{58}$ has a mutual ascending nodal distance of 1.35~au 
              at 339~au from the Sun. We estimate that the probability of finding such a pair by chance is 0.000205$\pm$0.000005. 
        \item The orientations in space of the orbits of the ETNOs show an asymmetric distribution of angular separations between perihelia 
              that we tentatively attribute to observational bias.
        \item Although most angular separations between orbital poles and perihelia are close to 30\degr, there are some pairs of ETNOs 
              with similar values of poles and perihelia. The strongest outlier is the pair 2013~FT$_{28}$--2015~KG$_{163}$. 
        \item The observed distributions of $\Omega$, $\omega$ and $\varpi=\Omega+\omega$ of the ETNOs appear to be compatible with those of 
              a uniform or random sample.
        \item The inclination distribution of the known ETNOs seems to follow the law discussed by \citet{2001AJ....121.2804B,
              2017AJ....154...65B}
     \end{enumerate}
     The statistical significance of our results is unexpectedly robust, the simple scattered disc model discussed in Section~6 is able to 
     reproduce relatively well the observed outlier limits for ${\Delta}_{+}$, $\alpha_q$ and $\alpha_p$ as well as the actual distributions 
     in ${\Delta}_{+}$ and $\alpha_p$; it fails, however, in reproducing the observed distributions of ${\Delta}_{-}$ and $\alpha_q$. 
     Observational bias may be regarded as the origin of the discrepancies in $\alpha_q$, but the ${\Delta}_{+}$--${\Delta}_{-}$ asymmetry 
     is probably not due to observational bias. Our results show that one may have uniformity in $\Omega$ and $\omega$ within the extreme 
     outer Solar system orbital parameter space as advocated by e.g. \citet{2017AJ....154...50S} or \citet{2020PSJ.....1...28B} or 
     \citet{2021PSJ.....2...59N} and still preserve tantalizing signals of present-day gravitational perturbations like the highly 
     statistically significant asymmetry between the shortest mutual ascending and descending nodal distances identified here or the 
     existence of at least one peculiar pair of ETNOs ---505478 and 2016~SG$_{58}$--- with a small value of the mutual nodal distance and 
     located beyond 300~au from the barycentre of the Solar system. 

  \section*{Acknowledgements}
     We thank the referee for her/his constructive, actionable, and insightful reports that included very helpful suggestions regarding the
     presentation of this paper and the interpretation of our results, J. Giorgini for extensive comments on JPL's SBDB and HORIZONS 
     systems, S.~J. Aarseth, J. de Le\'on, J. Licandro, A. Cabrera-Lavers, J.-M. Petit, M.~T. Bannister, D.~P. Whitmire, G. Carraro, E. 
     Costa, D. Fabrycky, A.~V. Tutukov, S. Mashchenko, S. Deen and J. Higley for comments on ETNOs, and A.~I. G\'omez de Castro for 
     providing access to computing facilities. This work was partially supported by the Spanish `Ministerio de Econom\'{\i}a y 
     Competitividad' (MINECO) under grant ESP2017-87813-R. In preparation of this paper, we made use of the NASA Astrophysics Data System 
     and the MPC data server.

  \section*{Data Availability}
     The data underlying this paper were accessed from JPL's SBDB (\href{https://ssd.jpl.nasa.gov/sbdb.cgi}{https://ssd.jpl.nasa.gov/sbdb.cgi}).
     The derived data generated in this research will be shared on reasonable request to the corresponding author.

  \appendix

  \section{Mutual nodal distances and uncertainty estimates}
     The mutual nodal distance between two Keplerian trajectories with a common focus can be written as (see eqs.~16 and 17 in 
     \citealt{2017CeMDA.129..329S}):
     \begin{equation}
        {\Delta}_{\pm} =  \frac{a_{2} \ (1 - e_{2}^2)}{1 \pm e_{2} \ \cos{\varpi_{2}}} - \frac{a_{1} \ (1 - e_{1}^2)}{1 \pm e_{1}
                           \ \cos{\varpi_{1}}} \,,
        \label{noddis}
     \end{equation}
     where for prograde orbits the `+' sign refers to the ascending node and the `$-$' sign to the descending one, and
     \begin{equation}
        \resizebox{0.9\hsize}{!}{$
        \cos{\varpi_{1}} =  \frac{-\cos{\omega_{1}} \ (\sin{i_{1}}\ \cos{i_{2}} - \cos{i_{1}}\ \sin{i_{2}}\ \cos{\Delta\Omega}) +
                                 \sin{\omega_{1}}\ \sin{i_{2}}\ \sin{\Delta\Omega}}
                                 {\sqrt{1 - (\cos{i_{2}}\ \cos{i_{1}} + \sin{i_{2}}\ \sin{i_{1}}\ \cos{\Delta\Omega})^2}}
                                 $}
        \label{cosvarpi1}
     \end{equation}
     and
     \begin{equation}
        \resizebox{0.9\hsize}{!}{$
        \cos{\varpi_{2}} =  \frac{\cos{\omega_{2}} \ (\sin{i_{2}}\ \cos{i_{1}} - \cos{i_{2}}\ \sin{i_{1}}\ \cos{\Delta\Omega}) +
                                \sin{\omega_{2}}\ \sin{i_{1}}\ \sin{\Delta\Omega}}
                                {\sqrt{1 - (\cos{i_{2}}\ \cos{i_{1}} + \sin{i_{2}}\ \sin{i_{1}}\ \cos{\Delta\Omega})^2}}
                                  \,,$}
        \label{cosvarpi2}
     \end{equation}
     with $\Delta\Omega = \Omega_{2} - \Omega_{1}$, and $a_j$, $e_j$, $i_j$, $\Omega_j$ and $\omega_j$ ($j=1, 2$), are the orbital elements 
     of the orbits involved. A small mutual nodal distance implies that in absence of protective mechanisms the objects may experience close
     flybys and, if the distance is small enough, even a collision. In order to obtain the actual distributions of ${\Delta}_{\pm}$, we 
     generated sets of orbital elements for the virtual ETNOs using data from Table~\ref{etnosB}. For example, the value of the semimajor 
     axis of a virtual ETNO was computed using the expression $a_{\rm v}=a_{\rm b}+\sigma_{a}\,r_{\rm i}$, where $a_{\rm b}$ is the 
     barycentric semimajor axis from Table~\ref{etnosB}, $\sigma_{a}$ is the standard deviation from Table~\ref{etnosB}, and $r_{\rm i}$ is 
     a (pseudo) random number with normal distribution computed using NumPy (van der Walt et al. \citeyear{2011CSE....13b..22V};
     \citealt{2020Natur.585..357H}). In order to calculate statistically relevant values of ${\Delta}_{\pm}$, we computed median and 16th 
     and 84th percentiles from a set of 10$^4$ pairs of virtual ETNOs for each actual pair from Table~\ref{etnosB}. Although the 
     distributions are in general not normal, in a normal distribution a value that is one standard deviation above the mean is equivalent 
     to the 84th percentile and a value that is one standard deviation below the mean is equivalent to the 16th percentile (see e.g. 
     \citealt{2012psa..book.....W}). By providing these values (standard deviation and the relevant percentiles), we wanted to quantify how 
     non-normal the distributions are for each pair. In some cases the resulting distribution is virtually normal, but in those with large 
     uncertainties, the skewness and tails could be significant. 

  \section{Poles, perihelia and uncertainty estimates}
     The ecliptic coordinates of the pole of an orbit are $(l_{\rm p}, b_{\rm p}) = (\Omega-90\degr, 90\degr-i)$ and those of the longitude 
     and latitude of an object at perihelion, $(l_q, b_q)$, are given by the expressions: $\tan{(l_q-\Omega)}=\tan\omega\,\cos{i}$ and 
     $\sin{b_q}=\sin\omega\,\sin{i}$ (see e.g. \citealt{1999ssd..book.....M}). The angular separations between orbital poles, 
     $\alpha_{\rm p}$, and perihelia, $\alpha_q$, are given by the expressions:
     \begin{equation}
        \cos{\alpha_{\rm p}} = \cos{b_{\rm p2}} \ \cos{b_{\rm p1}} * \cos{(l_{\rm p2} - l_{\rm p1})} + \sin{b_{\rm p2}} \ \sin{b_{\rm p1}} 
     \end{equation}
     and
     \begin{equation}
        \cos{\alpha_{q}} = \cos{b_{q2}} \ \cos{b_{q1}} * \cos{(l_{q2} - l_{q1})} + \sin{b_{q2}} \ \sin{b_{q1}} \,,
     \end{equation} 
     where the subscripts 1 and 2 refer to the members of the pair. As in the case of ${\Delta}_{\pm}$, we computed median and 16th and 
     84th percentiles of $\alpha_{\rm p}$ and $\alpha_q$ from a set of 10$^4$ pairs of virtual ETNOs for each actual pair from 
     Table~\ref{etnosB}. 

  \section{Brown distribution of inclinations}
\begin{figure}
       \centering
        \includegraphics[width=\linewidth]{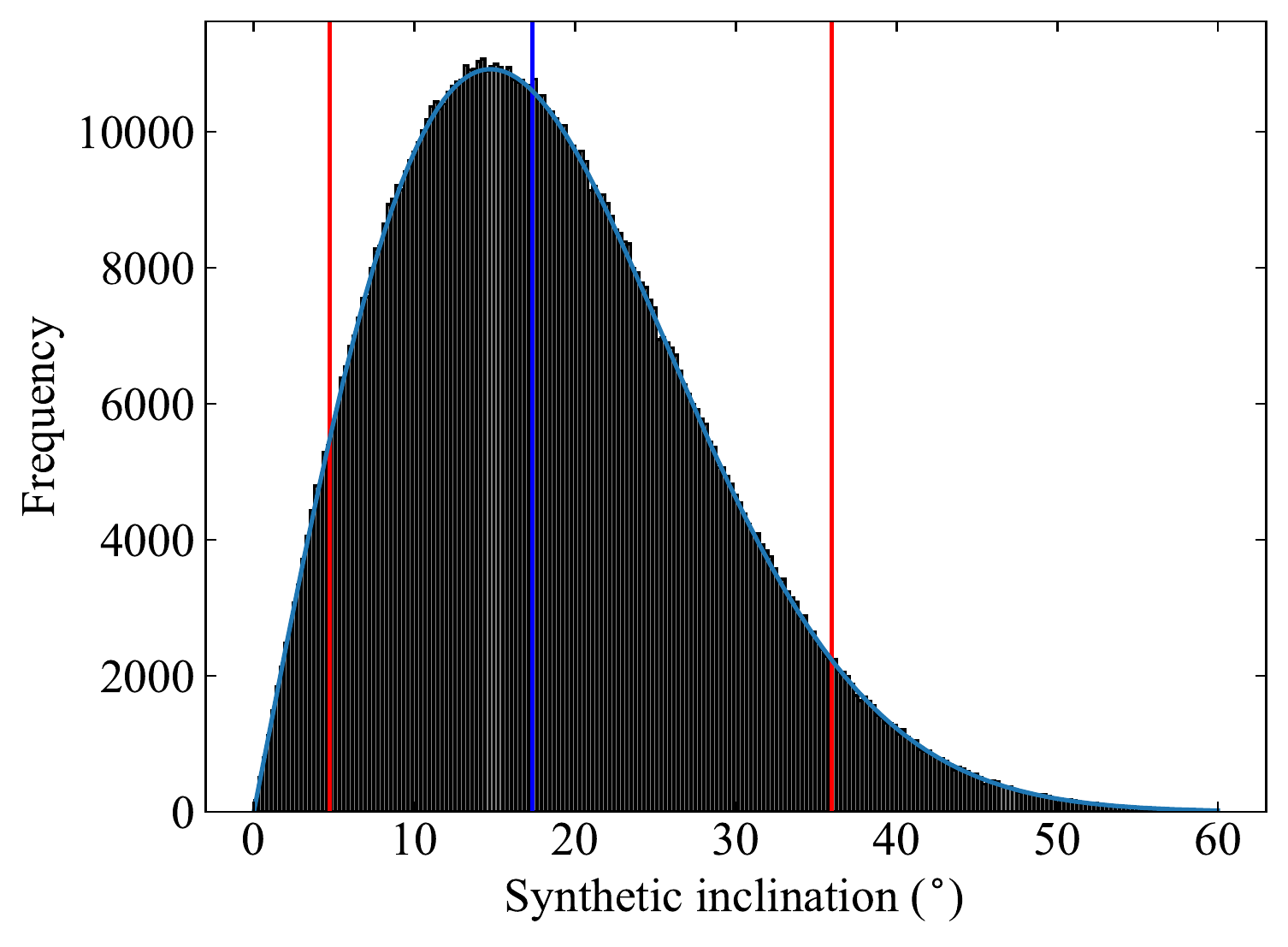}
        \caption{Distribution of inclinations consistent with the results in \citet{2001AJ....121.2804B}. The median is shown in blue and 
                 the 5th and 95th percentiles in red. The histogram corresponds to a sample of 10$^6$ values. The analytical curve is shown 
                 in grey.
                }
        \label{browni}
     \end{figure}

     \citet{2001AJ....121.2804B,2017AJ....154...65B} found that the distribution of inclinations for the `hot population' of the 
     trans-Neptunian or Kuiper belt can be described by the expression $N(i) \propto \sin{i}\ e^{\frac{-(i-\mu_{i})^2}{2\ \sigma_{i}^{2}}}$, 
     with $\mu_{i}=0\degr$ and $\sigma_{i}=15\degr$. In order to generate inclinations following the Brown distribution as continuous random 
     variables, we use the universal von Neumann rejection method also known as rejection sampling \citep{vonNeumann1951} implemented as a 
     Python function with $\sigma_{i}=14.9\degr$:  
     \begin{verbatim}
def browni(): # rejection sampling, 
              # von Neumann (1951)
    imin =  0.0    # minimum inclination
    imax = 60.0    # maximum inclination

    fmax =  0.1559783481171226 # maximum f(i)

    a = True
    while a:
       x = random.uniform(imin,imax)
       y = random.uniform(0,fmax)

       # probability density function, Brown (2017)
       f = math.sin(math.radians(x)) / 
           math.exp(x * x * 0.5 / 14.9 / 14.9)

       if y <= f:
          a = False

    return x
     \end{verbatim}
     The histogram of a set of 10$^6$ values generated with this function is shown in Fig.~\ref{browni}.

     The other orbital elements of the virtual TNOs in Section~6.2 are generated using the expressions:
     \begin{equation}
        \begin{aligned}
           a_{\rm v} & = 150 + 850\,r_{1} \\
           q_{\rm v} & = 30 + 70\,r_{2} \\
           e_{\rm v} & = 1.0 - q_{\rm v} / a_{\rm v} \\
           \Omega_{\rm v} & = 360\,r_{3} \\
           \omega_{\rm v} & = 360\,r_{4} \,,
           \label{virtual}
        \end{aligned}
     \end{equation}
     where $r_j$ with $j=1, 4$, are random numbers in the interval (0, 1) with a uniform distribution.
%
%
%
%
   
  \bsp
  \label{lastpage}
\end{document}